\documentstyle{mn}

\newcommand{\ie}{{\it i.e.\,}}

\newcommand{\cf}{{\it c.f.\,}}
\newcommand{\etal}{{\it et al.}}

\newcommand{\kms}{\mbox{${\,\rm km~s}^{-1}$}\,}
\newcommand{\kev}{\,\mbox{${\,\rm keV}$}}
\newcommand{\lsun}{\mbox{\,$L_\odot$\,}}

\newcommand{\msun}{\mbox{\,$M_\odot$\,}}
\newcommand{\msunyr}{\mbox{\,$M_\odot$\ {\rm yr}$^{-1}$\,}}
\newcommand{\ergs}{\mbox{${\,\rm erg~s}^{-1}$\,}}

\newcommand{\gmcmc}{\mbox{${\,\rm gm\ cm}^{-3}$\,}}

\newcommand{\surfb}{\mbox{${\,\rm erg~cm}^{-2} {\rm s}^{-1} {\rm arcmin}^{-2}$\,}}

\newcommand{\rosat}{\mbox{\sl ROSAT}}
\newcommand{\einstein}{\mbox{\sl EINSTEIN}}
\newcommand{\axaf}{\mbox{\sl Chandra}}
\newcommand{\xmm}{\mbox{\sl XMM}}

\title[Wakes and Tails Around Galaxies in Clusters]{Galaxies in
Clusters: The Observational Characteristics of Bow-Shocks, Wakes and Tails}

\author[I.\,R. Stevens, D.\,M. Acreman \& T.\,J. Ponman]{Ian R. Stevens,
David M. Acreman \& Trevor J. Ponman\\ 
School of Physics and Astronomy, University of Birmingham, 
Edgbaston, Birmingham, B15 2TT, UK}

\date{  Accepted ..............................; 
        Received ..............................; 
in original form ..............................}

\begin{document}

\maketitle

\begin{abstract}

The dynamical signatures of the interaction between galaxies in clusters
and the intracluster medium (ICM) can potentially yield significant
information about the structure and dynamical history of clusters. To 
develop our understanding of this phenomenon we present results from
numerical modelling of the galaxy/ICM interaction, as the galaxy moves
through the cluster. The simulations have been performed for a broad range,
of ICM temperatures ($kT_{cl}=1$, 4 and $8\kev$),
representative of poor clusters or groups through to rich clusters.

There are several dynamical features that can be identified in these
simulations; for supersonic galaxy motion, a leading bow-shock is
present, and also a weak gravitationally focussed wake or tail behind
the galaxy (analogous to Bondi-Hoyle accretion). For galaxies with higher
mass-replenishment rates and a denser interstellar medium (ISM), the
dominant feature is a dense ram-pressure stripped tail. In line with
other simulations, we find that the ICM/galaxy ISM interaction can
result in complex time-dependent dynamics, with ram-pressure stripping
occurring in an episodic manner.

In order to facilitate this comparison between the observational
consequences of dynamical studies and X-ray observations we have
calculated synthetic X-ray flux and hardness maps from these
simulations. These calculations predict that the ram-pressure stripped
tail will usually be the most visible feature, though in nearby galaxies
the  bow-shock preceeding the galaxy should also be apparent in deeper
X-ray observations. We briefly discuss these results and compare with
X-ray  observations of galaxies where there is evidence of such interactions.

\end{abstract}


\section{Introduction}
\label{sec1}
Most galaxies are not isolated (Tully 1987) but are members of groups
or clusters which are gravitationally bound, though not necessarily 
virialised. Galaxies in such systems may experience a number of 
types of interaction with their environment: galaxy-galaxy encounters,
which may trigger starbursts and can culminate in galaxy merging
(Schweizer 1982), interaction with the potential of the system as a whole,
which may for example, result in tidal stripping of loosely bound
material from galaxy outskirts (Mamon 1987), and finally hydrodynamical
interaction between the interstellar medium (ISM) in a galaxy, and
any intergalactic gas trapped in the potential well of the galaxy system.

It is the last of these processes which is the subject of the present
study. Interaction with the surrounding intracluster
medium (ICM) may have a significant impact on the gas within a galaxy
whenever it sweeps through a substantial mass of hot gas in its motion
within a group or cluster. In practice, the hot ICM dominates the baryon
content of galaxy clusters (White \& Fabian 1995) and is also a major
baryonic component within many galaxy groups 
(Mulchaey \etal\ 1996, Ponman \etal\ 1996).

The interaction between galaxies and the ICM should generate
characteristic structures: at low galaxy velocities it will produce only a
mild displacement of the outer regions of any gaseous galaxy halo, at
higher velocities a bow shock should be generated ahead of the galaxy,
and a wake of enhanced gas density may trail behind it.

These processes are important for three main reasons: first, gas may be
removed from and/or accreted by the galaxy in such interactions with its
gaseous environment, with important implications for subsequent galaxy
evolution. Second, such processes will be responsible for some or all
of the metals which are observed today within the ICM in clusters and
groups, and may also affect the energetics of the ICM (Ponman, Cannon \&
Navarro 1999) -- as such they are important for an understanding of the
structure and evolution of galaxy systems. Finally, structures such as
galaxy wakes provide information about the direction of motion of
galaxies on the plane of the sky, which has in general been unavailable
previously (Merrifield 1998). Given a good understanding of the
astrophysics of the interaction, one may also be able to deduce the
magnitude of the galaxy's total velocity (allowing a full dynamical
solution) and also to constrain the total mass of galaxies, including any
dark halo.

Given the velocities of galaxies in groups and clusters, the temperature
of the gas in shocks and wakes will be $\sim 10^7$\,K, so the
observational evidence of such processes is best sought at X-ray
energies. To date shock and wake structures have only been observed in a
few instances (detailed below), due to the limited sensitivity of
existing X-ray telescopes. However, the improvements available from the
upcoming X-ray observatories, \axaf\ and \xmm\, will enable detailed
studies of structures around galaxies within clusters to be conducted.
The aim of the present paper is to prepare for this, by exploring the
astrophysics of the galaxy-ICM interaction using numerical hydrodynamic
simulations, and in particular to investigate the observable properties
of such interactions in the X-ray band.

The astrophysics is quite complex -- several different mechanisms may be
at work in any particular galaxy, and these will potentially have
different signatures. Examples of such mechanisms include: i) a
bow-shock, if the galaxy is moving supersonically through the ICM then
there is the likelihood of a bow-shock preceeding the galaxy; ii) a
gravitationally focussed wake, the mass and motion of the galaxy will
tend to concentrate material behind it, in an analogous manner to
Bondi-Hoyle accretion in binary systems (\cf Hunt 1971); iii) ram
pressure stripping of gas in the galaxy by the ICM will also result in a
tail of enhanced density behind the galaxy. This feature, in the case of
M86, has also been referred to as an X-ray plume (Rangarajan \etal\ 1995).

Examples of these features can be seen later in simulations of the
galaxy/ICM interaction. While a bow-shock will be a distinct and
recognisable feature, distinguishing between wakes (formed by
gravitational focussing) and tails formed by stripping might be difficult
on morphological grounds. In this paper we identify spectral as well as
morphological signatures that can be used as a diagnostic of the physical
mechanisms at work.

Currently there is only a limited amount of evidence of bow-shocks, wakes
and tails around galaxies, observed primarily in relatively local
systems. Some examples are:

\begin{enumerate}

\item White \etal\ (1991) reported on \einstein\ observations
of the Virgo galaxy M86, finding evidence for an X-ray plume, caused by
ram pressure stripping as the galaxy moves through the cluster ICM.
Rangarajan \etal\ (1995) analysed \rosat\ observations of M86, finding
that the X-ray plume, which they also interpret as being due to
ram-pressure stripping, has a higher metallicity than either the
cluster or M86 itself. The plume temperature is $kT=0.8\kev$, similar to
the galaxy emission from M86, but substantially lower than the cluster.

\item David \etal\ (1994) reported on \rosat\ observations of a linear
structure behind the dominant galaxy in the NGC\,5044 group. they
interpreted these results as being due to a \lq\lq cooling 
wake\rq\rq\ behind
the galaxy, caused by motion of the galaxy with respect to the hot gas
which follows the potential of the galaxy group.

\item Irwin \& Sarazin (1996) reported on \rosat\ HRI observations of the
Virgo cluster elliptical galaxy NGC\,4472, which shows a distorted
X-ray morphology.  Irwin \& Sarazin (1996) suggested that this was 
strong evidence for a bow-shock like structure around the galaxy.

\item Sakelliou, Merrifield \& McHardy (1996) reported on \rosat\
results on observations of the wide-angled tail galaxy 4C\ 34.16
which lies in a poor cluster. Radio maps show two tails bent into a wide
C-shape, and X-ray observations show an enhancement between the tails.
Sakelliou \etal\ (1996) interpret this as due to motion of the galaxy
through the ICM, and that this motion is responsible for both the jet
bending and the X-ray wake trailing the galaxy.

\item The Fornax elliptical galaxy NGC\,1404 shows an X-ray tail pointing
away from the cluster centre, indicating that galaxy is infalling (Jones
\etal\ 1997).

\item \rosat\ observations of the Coma cluster found an elongated
X-ray structure around the giant elliptical galaxy NGC\,4839 (Dow \& White
1995). NGC\,4839 is probably the cD galaxy of a subcluster that is
currently infalling into the main cluster. The X-ray tail of NGC\,4839 is
pointing away from the main cluster centre, supporting the infall idea.

\item H$\alpha$ observations of the edge-on spiral galaxy NGC\,4388 by 
Veilleux \etal\ (1999) have found evidence of supersonic ram-pressure
stripping. NGC\,4388 is moving
at $\sim 1500\kms$ with respect to the ICM of the Virgo cluster, with a Mach
number of ${\cal M}=3$. The X-ray emission from NGC\,4388 is extended
in the plane of the galaxy (Matt \etal\ 1994) but there is no evidence 
of X-ray extensions
associated with ram-pressure stripping (though as Veilleux \etal\ (1999)
point out, the galaxy orientation is unfavourable).

\end{enumerate}

There have been a several papers considering the gas-dynamics of a galaxy
moving through a cluster since the issue was considered by Gunn \& Gott
(1972). Shaviv \& Salpeter (1982) reported on
simulations, concentrating on the role of galaxy motion in ram-pressure
stripping. Further simulations of the galaxy/ICM interaction have been
presented by Takeda, Nulsen \& Fabian (1984), who presented results of
the dynamics of a galaxy moving radially through a cluster, suffering
periodic stripping in the higher density regions near the centre of a
cluster. Other studies of ram-pressure stripping have been presented by
Nulsen (1982) and Portnoy, Pistinner \& Shaviv (1993), as well as the work of
Abadi, Moore \& Bower (1999), which concentrates on the effect of ram-pressure
stripping in clusters on the truncation of disks of spiral galaxies. 
The two most
relevant papers to this work are those by Gaetz, Salpeter \& Shaviv
(1987) and Balsara, Livio \& O'Dea (1994). Gaetz \etal\ (1987) presented a
comprehensive study of ram-pressure stripping from galaxies, and
calculated several important parameters affecting the efficiency of
stripping. We shall use some of the parameters defined 
by Gaetz \etal\ (1987) in this paper. Balsara \etal\ (1994) present
further calculations, using more sophisticated numerical
techniques. Importantly, they found that the stripping process was much
more complicated than seen in the calculations of Gaetz \etal\
(1987). When there was significant mass-replenishment of the ISM  within
the galaxy the resulting dynamical structure was complex, with material
being stripped from the galaxy in long streamers from the edge of the
galaxy, with the stripping happening is a discontinuous, episodic manner.

The calculations presented here represent an improvement over those
of Gaetz \etal\ (1987) and Balsara \etal\ (1994) for the following reasons.
The Gaetz \etal\ (1987) paper used a rather old numerical hydrodynamics
code, which has substantial numerical viscosity. This has the effect of
smearing out smaller scale features found in runs using more
recent numerical hydrodynamic computer codes, such as the Piecewise
Parabolic Method (PPM). This is important as in many runs the stripping
process occurs in discrete events, with material being stripped of the
edges of the galaxy. It is likely that in the Gaetz \etal\ (1987) models
these processes were not be resolved. Balsara \etal\ (1994)
used a more sophisticated code (PPM - essentially the same as used in
this paper). However, Balsara \etal\ (1994)
used a large value for the mass-replenishment rate ($\dot M_{rep}$). A
large value of $\dot M_{rep}$ has the advantage of highlighting the
complex dynamics of the galaxy/ICM interaction, but care has to be
taken not to make it unrealistically large (see Section~\ref{sec2p3p3}).

There are two further areas where this work is a significant improvement.
First, from the simulations we calculate synthetic flux and hardness
maps, as well as surface brightness profiles. These synthetic results can
then be compared directly to actual observations of structures around
galaxies in clusters. Such an approach will be necessary to fully
understand the observed structures. Second, we perform a broad ranging
parameter study, for parameters appropriate for a range of clusters,
ranging from cool clusters or groups (with an ICM temperature of $kT_{cl}\sim
1\kev$), up to hot clusters ($kT_{cl}\sim 8\kev$).

The main purposes of this paper are twofold. First, to identify
observational signatures of the galaxy/ICM interaction, which can be used
in future detailed studies of individual objects, and second, to identify
the circumstances in which galaxy/ICM interactions are most likely to be
visible.

The paper is organised as follows: in Section~\ref{sec2} we discuss the
hydrodynamical model used to investigate the phenomenon, including the
model for the galaxy and the parameters used in the simulations. In
Section~\ref{sec3} results from the simulations are presented and compared
with analytic models of the galaxy/ICM interaction, in
Section~\ref{sec4} the observable signatures of these
simulations are discussed, and in Section~\ref{sec5}
we discuss these results in the context of observations and conclude.

\begin{table*}
\caption{The assumed parameters for the model clusters. The various
symbols are defined in more depth in Section~\ref{sec2p2}.}
\begin{tabular}{llccc} \smallskip
Parameter & Symbol (Units) & Cool Cluster & Intermediate Cluster & 
Hot Cluster\\
Cluster X-ray luminosity & $L_x$ (\ergs)  & $6.1\times 10^{42}$ 
& $2.8\times 10^{44}$ & $1.9\times 10^{45}$\\
Cluster core radius      & $R_{core}$ (kpc)  & 250 & 250 & 250\\
Velocity dispersion      & $\sigma_{cl}$ (\kms) & 371 & 795 & 1164\\
Cluster temperature  & $kT_{cl}$ (keV) & 1.0 & 4.0 & 8.0\\
Cluster sound speed  & $a_{cl}$ (\kms)    & 520  & 1040 & 1460\\
Cluster density at galaxy radius$^1$ & $\rho_{cl}(R_{core})$ (\gmcmc) & 
$3.8\times 10^{-28}$ & $2.5\times 10^{-27}$ & $5.8\times 10^{-27}$\\
Central cluster surface brightness$^2$ & $S_{0}$ (\surfb) & 
$1.1\times 10^{-14}$ & $5.1\times 10^{-13}$ & $3.4\times 10^{-12}$\\
\end{tabular}
\begin{flushleft}
Notes for Table~1:\\
$^1$ In all simulations the galaxy is assumed to be at a radius of 
$r=R_{core}$ from the cluster centre.\\
$^2$ The quoted surface brightness is calculated assuming the 
cluster bolometric X-ray luminosity. The surface brightness in the 
\rosat\ waveband will be correspondingly lower.
\end{flushleft}
\label{tab:x1}
\end{table*}

\section{The Numerical Model}
\label{sec2}

\subsection{Introduction}
\label{sec2p1}

To model the gas dynamics in and around a galaxy moving 
within a cluster
we use the 2-D hydrodynamics code VH-1, which has been used 
extensively elsewhere in different contexts 
(Blondin \etal\ 1990; Stevens \etal\ 1992). The code uses the PPM
algorithm and is good at resolving shocks in complex flows. It is
basically very similar to that used by Balsara \etal\ (1994). The flow
is assumed to be cylindrically symmetric.

In these simulations the gravitational effect of the galaxy on
the ICM are included, as well as replenishment of the galaxy ISM by
stellar populations. The parameters for both the galaxy and the
clusters they are embedded in are described below.

\subsection{The Model Galaxy Cluster}
\label{sec2p2}

One of the main goals of this paper is to investigate the expected
observational visibility of the galaxy/ICM interaction region. To do
this we embed the results from the hydrodynamic simulations into a model
cluster. We assume a spherical, isothermal cluster with a temperature
$T_{cl}$ and cluster sound speed $a_{cl}$ (calculated assuming a mean
mass per particle of $\bar\mu=10^{-24}$\,gm). Simulations for three different
values of $T_{cl}$; a) $T_{cl}=1\kev$ have been performed, representative of a
poor cluster, such as Fornax (\cf 4C\,34.16, Sakelliou \etal\ 1996), b)
$T_{cl}=4\kev$, representative of a more substantial cluster, such as
Abell~1060 or AWM~7 (both of which have temperatures $\sim4\kev$;
Loewenstein \& Mushotzky 1996), and c) $T_{cl}=8\kev$, indicative of a
rich cluster such as Coma.

We assume the following scaling relationships for the cluster X-ray
luminosity ($L_x$) and velocity dispersion ($\sigma_{cl}$, White, Jones
\& Forman 1997):

\begin{equation}
L_x (\ergs) = 6.1\times 10^{42} \left(kT_{cl}\right)^{2.77}
\label{eq1}
\end{equation}

\begin{equation}
\sigma_{cl} (\kms)= 371 \left(kT_{cl}\right)^{0.55}
\label{eq2}
\end{equation}
with $kT_{cl}$ in \kev. For the model clusters a $\beta=2/3$ mass
distribution has been assumed (see below). The X-ray luminosity given by
eqn.\,(\ref{eq1}) is the total bolometric X-ray luminosity, rather than the
\rosat\ luminosity. A cluster core radius of $R_{core}=250$\,kpc and a
distance of $D=50$\,Mpc is assumed for all clusters (\cf Jones \& Forman 1984).

Future studies of individual galaxies in clusters will be able to 
use values of $L_x$ (and $\beta$ and $\sigma_{cl}$ etc) appropriate for the
cluster in question. However, these scaling relationships are adequate
for a general study of the observational characteristics
of structures around galaxies. For simplicity, the galaxy is assumed to
be moving through material of constant density, at a distance of
$r=R_{core}$ from the cluster centre. 

For $\beta=2/3$, the surface brightness distribution $S(w)$ of the model
cluster as a function of projected distance from the cluster centre $w$ is

\begin{equation}
S(w) =  \frac{S_0}{\left(1+(w/R_{core})^2\right)^{3/2}}\ ,
\end{equation}
where $R_{core}$ is the cluster core radius, and $S_0$ is the
central surface brightness. This profile integrates to give the
a total X-ray luminosity of $L_x = 8\pi^2 S_0 R_{core}^2$. Also, for
$\beta=2/3$, the gas density profile within the cluster $\rho_cl(r)$
varies as
\begin{equation}
\rho_{cl}(r) =  \frac{\rho_{cl}(0)}{\left(1+(r/R_{core})^2\right)}
\end{equation}
with $\rho_{cl}(0)$ the cluster central gas density. At $r=R_{core}$ the gas
density $\rho_{cl}(R_{core})=\rho_{cl}(0)/2$. Details of 
the parameters for the model clusters
are given in Table~\ref{tab:x1}.

\subsection{The Model Galaxy}
\label{sec2p3}

\subsubsection{Mass Distribution}
\label{sec2p3p1}
The parameters for the model galaxy used here are similar to those of 
Gaetz \etal\ (1987). The galaxy is assumed to be spherical, with the
mass in the galaxy dominated by stars. The stellar density
distribution  within the galaxy is given by
\begin{equation}
\rho_\ast(r)= \frac{\rho_\ast(0)}{\left(1+(r/R_c)^2\right)} \ ,
\end{equation}
with $R_c$ the core radius of the mass distribution ($R_c=1.5$\,kpc in 
these simulations), and $\rho_\ast(0)$ is the
central density in the galaxy. We assume that the mass distribution of
the galaxy extends out to a radius $R_H=32.2$\,kpc
(Balsara \etal\ 1994). The total galaxy mass, $M_{gal}$, is then

\begin{equation}
M_{gal}=4\pi \rho_\ast(0) R_c^3\left[\frac{R_H}{R_c}-
\tan^{-1}\left(\frac{R_H}{R_c}\right)\right]\ .
\end{equation}
In these simulations $M_{gal}=1.2\times 10^{12}\msun$, which then
specifies the galaxy central stellar density $\rho_\ast(0)$. The assumed
parameters for the model galaxy  are summarised in Table~\ref{tab:x2}.

\subsubsection{Galaxy Velocity}
\label{sec2p3p2}

For a galaxy cluster with a one-dimensional 
velocity dispersion of $\sigma_{cl}$ the
mean galaxy velocity will be $V_{mean}=\sqrt{3}\sigma_{cl}$.
The model galaxy is assumed to be moving at a velocity $V_{gal}$
through the ICM, and for each of the three model clusters 
simulations with two values of $V_{gal}$, have been run, 
with $V_{gal}=1.0$ or $1.5 V_{mean}$.
The relationship between $\sigma_{cl}$ and cluster temperature
(eqn.\,\ref{eq2}) means that $V_{gal}\propto kT_{cl}^{0.55}$.

For these values of $V_{gal}$ all the model galaxies are moving
supersonically, with Mach numbers ${\cal M}=1.2-2.1$. This range of
values should be representative, though some galaxies of interest have
even higher Mach numbers (c.f. NGC\,4388, Veilleux \etal\ 1999).
The steeper dependence of $V_{gal}$ on $T_{cl}$ compared to the sound speed,
means that the galaxies in clusters with higher $kT_{cl}$ have marginally
higher Mach numbers. Details of the velocities for the different models
are in Table~\ref{tab:x3}.

\begin{table*}
\caption{Parameters for the model galaxy used in all the
simulations. The symbols used are defined in more depth in 
Section~\ref{sec2p3}.}
\begin{tabular}{lll} \smallskip
Parameter                 &   Symbol  & Value  \\
Galaxy mass               & $M_{gal}$ & $1.2\times 10^{12}\msun$ \\
Galaxy central stellar density & $\rho_\ast(0)$ & $9.48\times 10^{-23}\gmcmc$ \\
Galaxy core radius        & $R_c$        & 1.5\,kpc \\
Galaxy outer radius       & $R_H$        & 32.2\,kpc \\
Mass replenishment half-radius  & $R_0$  & 8.7\,kpc \\
Mass replenishment outer radius & $R_{rep}$ & 15.5\,kpc \\
Mass replenishment temperature  & $T_{rep}$ & $1.87 \times 10^{6}$\,K \\
Star formation density threshold & $\rho_{sf}$& $2.34\times 10^{-24}\gmcmc$ \\
Star formation temperature threshold & $T_{sf}$ & $2.0 \times 10^{4}$\,K \\
Star formation timescale   & $\tau_{sf}$   & $3.25 \times 10^{7}$\,yr \\
\end{tabular}
\label{tab:x2}
\end{table*}

\subsubsection{Mass replenishment within the galaxy}
\label{sec2p3p3}

In galaxies a combination of stellar mass loss and supernovae act to
continuously replenish the gas within the galaxy. This
mass-replenishment will affect the flow around the galaxy in several
ways. First, it provides material that can be continuously stripped and
second, it acts as barrier (if dense enough) to the ICM, resulting in
the formation of a bow-shock preceeding the galaxy. In these simulations
the radial distribution of the mass replenishment is assumed to have
basically the same form as that of the stellar distribution, that is

\begin{equation}
\dot \rho_{rep}(r)= \frac{\dot\rho_{rep}(0)}{\left(1+(r/R_c)^2\right)} 
\end{equation}
for $r\leq R_{rep}$, where $R_{rep}$ is the outer radius for 
mass-replenishment. The characteristic radius $R_0$ is the mass
replenishment half-radius - half of the mass-replenishment occurs inside
a radius $R_0$. 
This prescription is equivalent to \lq\lq Case 1\rq\rq\ in Gaetz \etal\ (1987).

This expression can be integrated 
to give an expression for the total mass-injection rate within
the galaxy 

\begin{equation}
\dot M_{rep}= 4\pi {\dot \rho_{rep}(0)} R_c^3\left[\frac{R_{rep}}{R_c}-
\tan^{-1}\left(\frac{R_{rep}}{R_c}\right)\right] \ .
\end{equation}

The value of $\dot\rho_{rep}(0)$ is determined by the assumed total rate
of mass injection into the galaxy $\dot M_{rep}$. Assuming a galactic
mass/light ratio of 20 (in solar units) gives $L=5\times 10^{10}\lsun$
for the model galaxy. This mass/light ratio implies a modest dark matter
halo. Faber \& Gallagher (1976) estimated that the rate of mass-replenishment
into an elliptical galaxy was $0.015\msunyr \times (10^9\lsun)$, so the
expected mass-replenishment rate for a $1.2\times 10^{12}\msun$
galaxy is $\sim 1\msunyr$. As discussed by Gaetz \etal\ (1987), the
relationship of Faber \& Gallagher (1976) was estimated assuming a 
single coeval populations of stars, with a current main sequence turn-off
mass of $1\msun$. Younger populations of stars will enhance the
mass-replenishment rate, and so the value of $1\msunyr$ should be
considered as a lower limit. The results of Mathews (1989) suggest a 
value of $\dot M_{rep} = 5\msunyr$ as a reasonable
upper limit, when there is mass-replenishment from a younger 
stellar population. Consequently, two sets of simulations
with $\dot M_{rep}=1\msunyr$ and $5\msunyr$ have been calculated. 
The temperature of the replenished gas is assumed to be
$T_{rep}=1.87\times 10^6$\,K, the same as in Balsara \etal\ (1994) and
Gaetz \etal\ (1987).

We note that Balsara \etal\ (1994) used a rather high value of the
mass-replenishment rate ($\sim 15\msunyr$). Such a large value leads
automatically to the formation of dense structures in the centre of the
galaxy. However, even with lower
values of $\dot M_{rep}$, under some circumstances, dense and complex 
structures are formed within the galaxy.

\subsubsection{Star-formation}
\label{sec2p3p4}
In several of the simulations presented in Section~\ref{sec3} a
considerable amount of cold, dense gas accumulates in the centre of the
galaxy. We include the possibility of star-formation in the
simulations, in an identical way to Balsara \etal\ (1994), which is
essentially as a mass-sink in the centre of the galaxy. No feedback
mechanisms (\ie the formation of a young stellar population leading to
an increase in the mass-replenishment rate) is included. Such
refinements could be easily included in more sophisticated models.

As implemented here the action of star-formation is to remove 
gas from the central regions of the galaxy. Material is assumed to form
stars or molecular clouds and effectively drops out of the hydrodynamic
equations. When star-formation does occur the
rate of change of gas density is given by;

\begin{eqnarray}
\dot\rho_{sf} & = & -\frac{\rho}{4\tau_{sf}}
\left(1+\tanh\left[\frac{\rho-\rho_{sf}}{0.1\rho_{sf}}\right]\right)
\nonumber \\
&& \times \left(1+\tanh\left[\frac{T_{sf}-T}{0.1T_{sf}} \right]\right)
\end{eqnarray}
In this equation $\rho_{sf}$ and  $T_{sf}$ are the effective density and
temperature thresholds above which and below which star-formation will
occur. The star-formation
time-scale $\tau_{sf}$ specifies the time-scale on which the
star-formation occurs. In these simulations $\rho_{sf} = 2.4\times
10^{-24}\gmcmc$, $T_{sf}=20000$K, and $\tau_{sf} = 3\times 10^7$ years.

In our simulations star-formation only occurs in a subset of the runs
(see Section~\ref{sec3}), and when it does occur it only happens in the
innermost regions of the galaxy. Star-formation, as implemented here,
does not have any substantial effect on the flow-properties outside of
these inner regions, but it does prevent the densities in the
central region of the galaxy growing, unchecked, to unphysical levels.

\begin{figure*}
\vspace{11.5cm}
\includegraphics{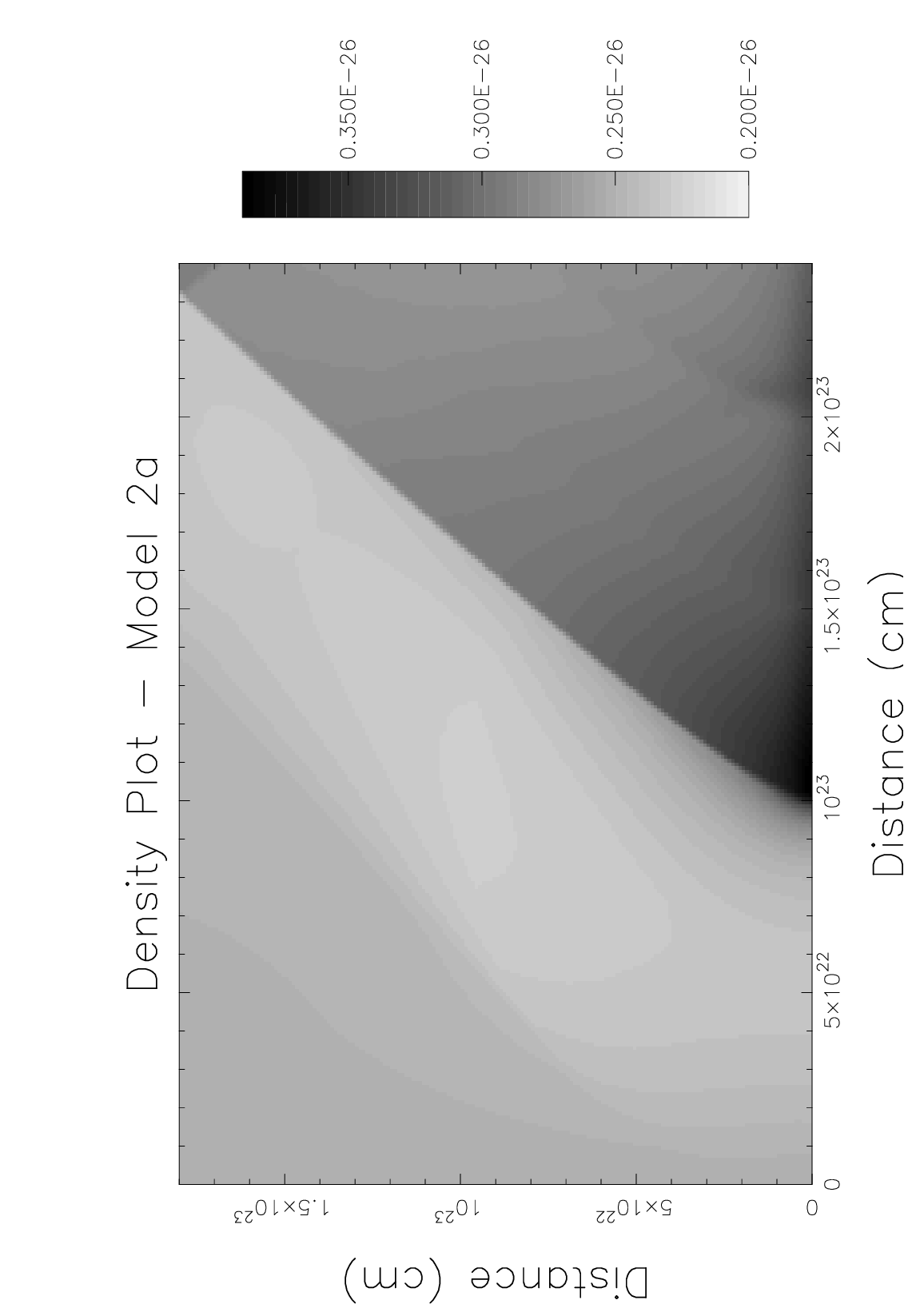}
\includegraphics{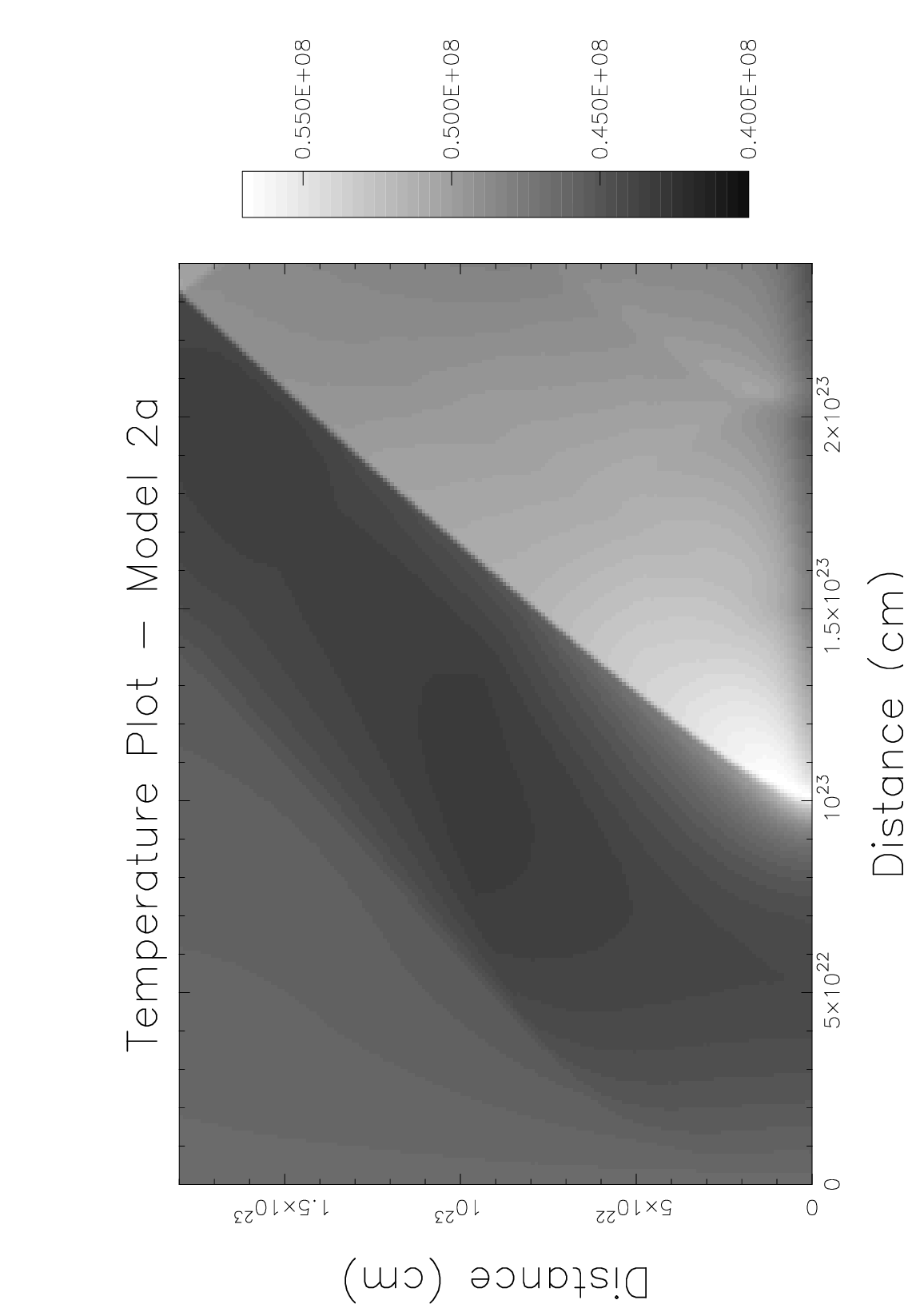}
\includegraphics{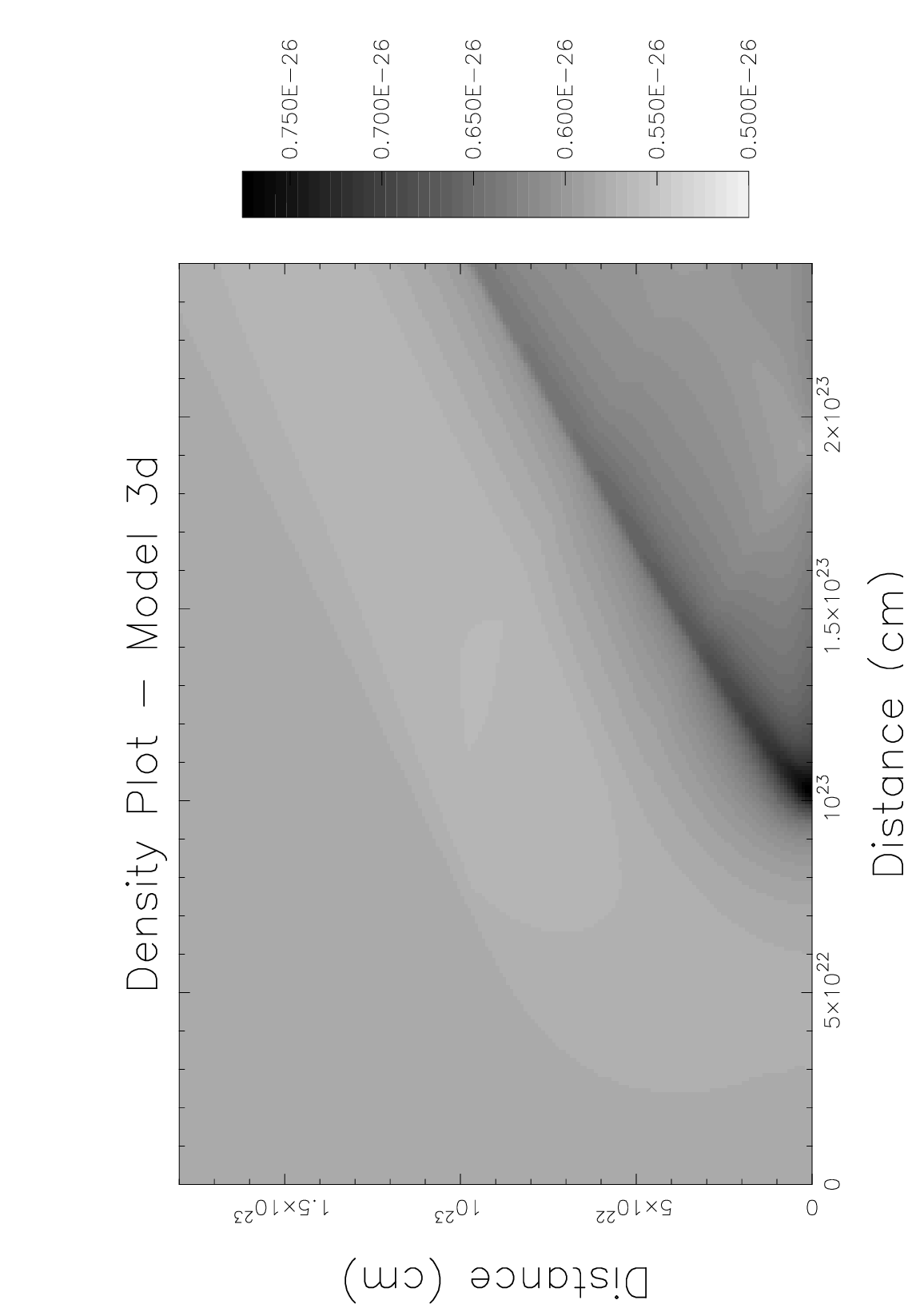}
\includegraphics{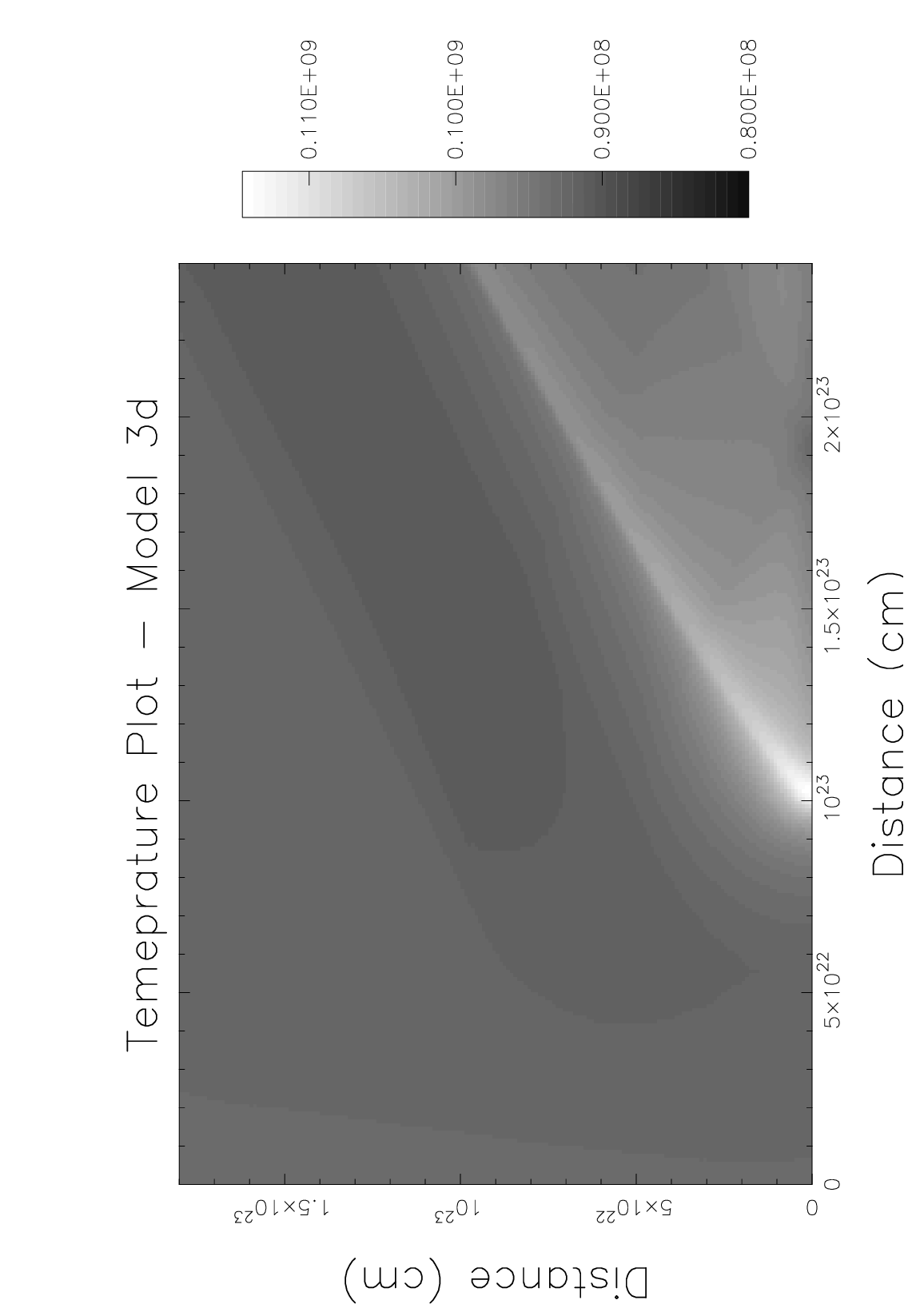}
\caption{The density and temperature structure for two simulations that
show the \lq\lq Efficient Stripping\rq\rq\ (ES) behaviour. Upper panels -
results for Model~2a (4\kev\ cluster - low velocity and low
mass-replenishment rate). Note the bow-shock structure, and also the
relatively small mass density/temperature perturbations associated with
the galaxy. Lower panels - results for Model~3d (8\kev\ cluster - high
velocity and high mass-replenishment). The bow shock is visible and has
a smaller opening angle due to the higher Mach number of the flow.}
\label{fig:x1}
\end{figure*}

\begin{figure*}
\vspace{11.5cm}
\includegraphics{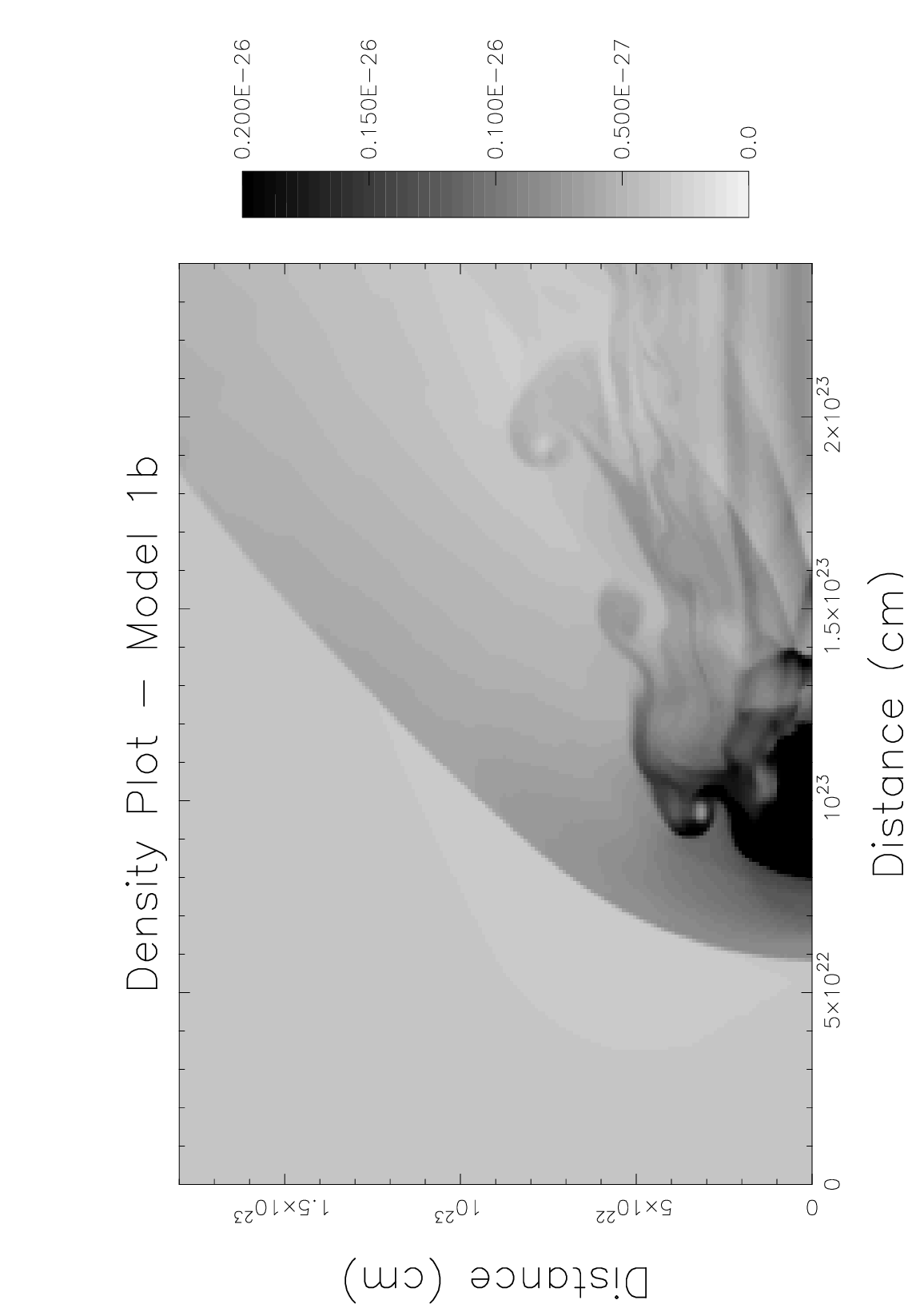}
\includegraphics{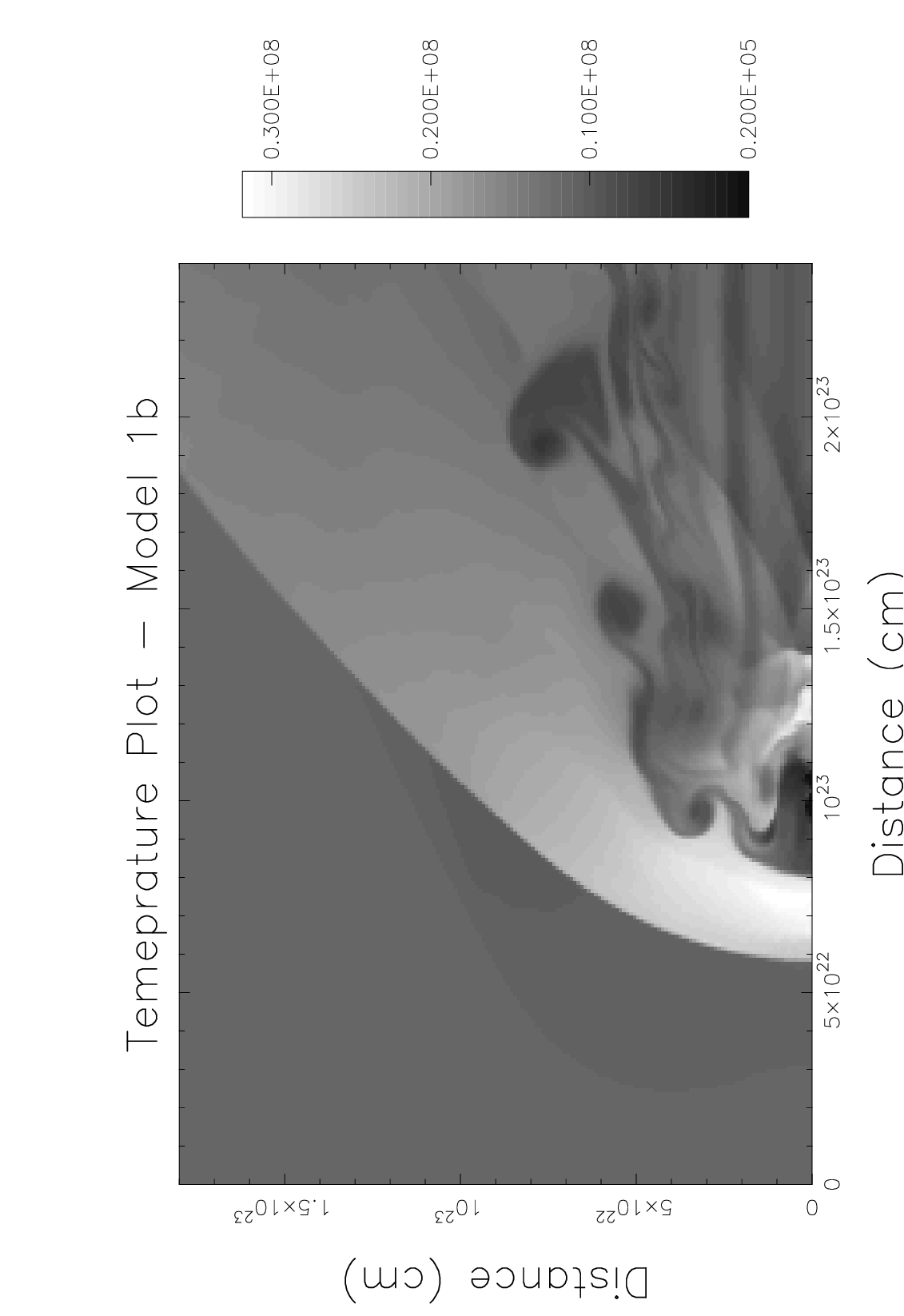}
\includegraphics{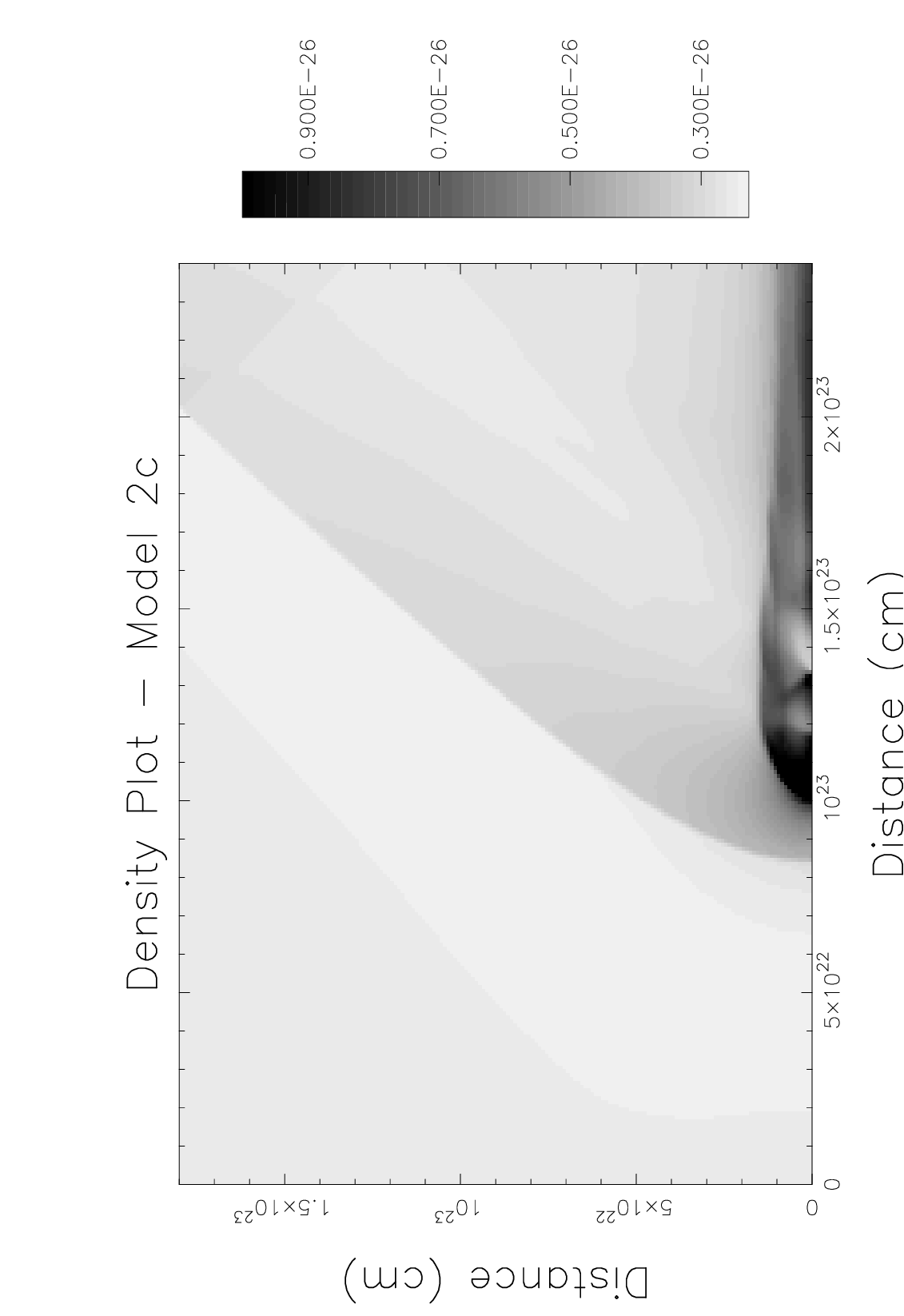}
\includegraphics{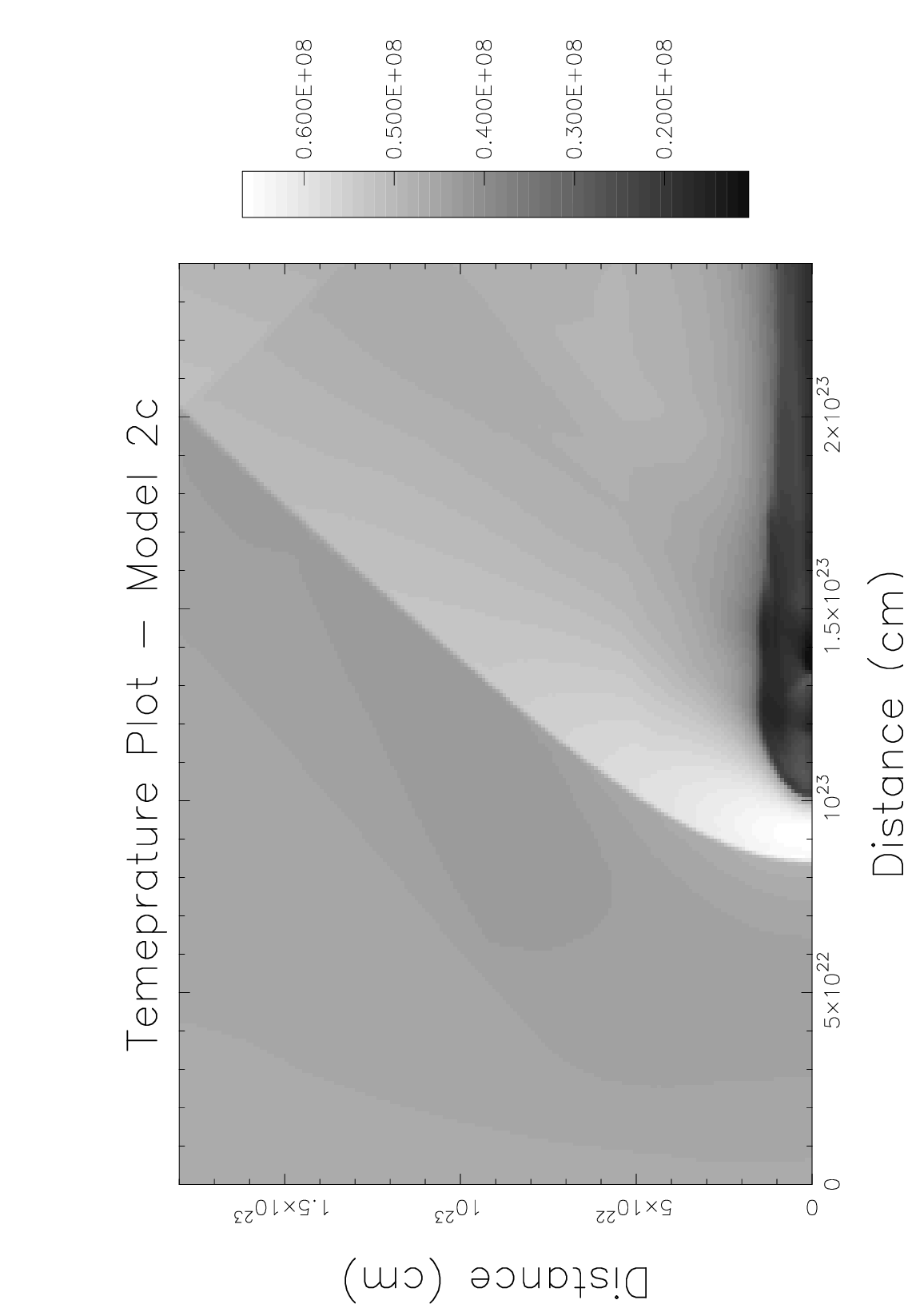}
\caption{The density and temperature structure for two simulations for the
\lq\lq Mass-Retention\rq\rq\ (MR) mode. Upper panels - results for Model~1b 
(1\kev\ cluster, low velocity, high mass-replenishment rate). Note the
strong density enhancements in the form of streamers in the tail region.
There is a considerable amount of cool and dense gas at the centre of
the galaxy. Lower panels - results for Model~2c (4\kev\ cluster, low
velocity, high mass-replenishment rate). The bow shock is again clearly
visible as well as a dense tail behind the galaxy caused by ram-pressure
stripping. The tail is not as over dense as for the case in the
top-panels, and is smaller in extent.}
\label{fig:x2}
\end{figure*}

\subsection{Description of Simulations}
\label{sec2p4}

A total of 12 simulations have been run, four each for the cool,
intermediate and hot clusters, with two values for the galaxy velocity and
mass-replenishment rate. Details of the parameters for each model are
given in Table~\ref{tab:x3}. The simulations were all run on a
$240\times 180$ 2-D cylindrical grid, with an  extent of $2.4\times
10^{23}$\,cm by $1.8\times 10^{23}$\,cm ($\sim 80\times 60$\,kpc). At an
assumed cluster distance of 50\,Mpc each cell corresponds to an angular extent 
of 1.3~arcsec. 

\subsection{Useful Parameters}
\label{sec2p5}

Gaetz \etal\ (1987) introduced some parameters and timescales that are
useful to characterise these simulations. The flow-crossing timescale is
defined as

\begin{equation}
\tau_{flow}= \frac{2R_0}{V_{gal}}\ .
\end{equation}
For the range of velocities considered
$\tau_{flow}$ is in the range $(0.6-2.6)\times 10^7$\,yr. The 
characteristic orbital timescale within the galaxy is defined as

\begin{equation}
\tau_{orb}= \frac{2\pi R_0}{v_{circ,0}}
\end{equation}
with

\begin{equation}
v_{circ,0}=\left(\frac{GM(R_0)}{R_0}\right)^{1/2}
\end{equation}
with $M(R_0)$ the total mass-contained within a radius $R_0$.
For the model galaxy $M(R_0)=2.6\times 10^{11}\msun$, so that 
$v_{circ,0}=360\kms$, and $\tau_{orb}=4.7\times 10^7$\,yr.

An important quantity in these simulations will
be the fraction of mass retained by the galaxy in the gas-phase and the
amount of gas ram-pressure stripped from the galaxy. Gaetz \etal\ (1987)
parameterised the fraction of mass ram-pressure stripped from the galaxy
using two parameters, $\nu$ and $\omega$, where

\begin{equation}
\nu=\frac{\pi R_0^2 \rho_{cl} V_{gal}}{\dot M_{rep}}\ ,
\end{equation}
with $\rho_{cl}$ the ICM density and
\begin{equation}
\omega=4\left(\frac{V_{gal}}{10^3\kms}\right)^2
\left(\frac{v_{esc,0}}{500\kms}\right)^{-2}\ .
\end{equation}
The escape velocity from the galaxy at a
radius $R_0$ is

\begin{equation}
v_{esc,0}=\left(2|\phi(R_0)-\phi(\infty)|\right)^{1/2}\ .
\end{equation}
with $\phi$ the gravitational potential.
For the parameters used here $v_{esc,0}=850\kms$ (Table~\ref{tab:x2}).

The parameter $\nu$ is the ratio of the external to internal mass-flow
rate into the galaxy, and $\omega$ a dimensionless measure of galaxy
velocity. With these two parameters Gaetz \etal\ (1987) defined a
dimensionless stripping parameter $\xi =\nu \omega^{0.7}$ that related
directly to the fraction of mass-stripped from the galaxy (their
Fig.~3). Consequently, the rate of ram-pressure stripping is $\propto 
\rho_{cl} V_{gal}^{2.4}$.

From the simulations we calculate the fraction  of the 
replenished gas that is retained by the galaxy ($\Phi$). We estimate
the mass of gas $M_{gas}$ within a radius $R_0$ at two times
$t_1$ and $t_2$, with $t_1$ and $t_2$ chosen to lie at times where the
solution has settled down to a steady state (or at least a quasi-steady
state - see Section~\ref{sec3p2}). We then define $\Phi$ as:

\begin{equation}
\Phi=\frac{\left[M_{gas}(t_2)- M_{gas}(t_1)\right] + \left[M_{sf}(t_2)-
M_{sf}(t_1)\right]}
{(t_2-t_1)\dot M_{rep}}\ ,
\label{eq16}
\end{equation}
where $M_{sf}(t)$ is the total mass removed from the system by
star-formation up to a time $t$. 
From their models, Gaetz \etal\ (1987) empirically found 
that $\Phi$ could be represented by the following expression

\begin{equation}
\Phi=\left(1+\left(\frac{\xi}{\xi_{1/2}}\right)^{1.23}\right)^{-1}
\label{eq17}
\end{equation}
with $\xi_{1/2}=0.6$ is the value of $\xi$ at which $\Phi=0.5$. Our
results for $\Phi$ will be discussed in Section~\ref{sec3}.

Other parameters from the simulations that will be useful for comparison
with observations are the emission weighted temperatures of specified
regions within the simulations. For all of the simulations we have
calculated the emission weighted temperature for two regions. The first 
covers the galaxy itself, and comprises a sphere, radius $R_0$, centred
on the galaxy ($\bar T_{gal}$). The second covers the tail region, and
is a cylinder, radius $R_0$, extending from a distance $R_0$ behind the
galaxy to the edge of the grid $(\bar T_{tail}$). The emission weighted
temperature $\bar T$ is defined as

\begin{equation}
\bar T = \frac{\sum_{cells} n^2 \Lambda(T) T }{\sum_{cells} n^2 \Lambda(T)}
\end{equation}
with $n$ the gas number density and $\Lambda(T)$ the emissivity of the
gas. The summation is carried out over all cells within the region of 
interest. The values of $\bar T_{gal}$ and $\bar T_{tail}$
for each of the simulations are shown in Table~\ref{tab:x3}, although it
should be noted that because some of the simulations never settle down
to a steady state, the values of $\bar T_{tail}$ and $\bar T_{gal}$ can 
vary with time.

\begin{table*}
\caption{The assumed and derived parameters for the simulations.}
\begin{tabular}{llcccccccl} \smallskip\\
&& $V_{gal}$ & $\dot M_{rep}$ & ${\cal M}$ & $\xi$ &
$\Phi$ & $\bar T_{gal}$ & $\bar T_{tail}$ & Mode \\
&  & (\kms) &  (\msunyr)  &  & & & (K)  & (K)  &   \\
&  & (1) & (2) & (3) & (4) & (5) & (6) & (7) & (8) \\
& & &&    &      & &      & &   \\
\multicolumn{8}{l}{Cool cluster models ($kT_{cl}=1\kev=1.2\times
10^7$K)} && \\
& Model 1a  & 640 & 1.0 & 1.2 & 0.59 & 0.33 & $1.2\times 10^5$ &
$1.1\times 10^7$ & MR \\
& Model 1b  & 960 & 1.0 & 1.9 & 1.6 & 0.31 & $3.5\times 10^4$ & 
$1.1\times 10^7$   & MR \\
& Model 1c  & 640 & 5.0 & 1.2 & 0.12 & 0.89 & $2.4\times 10^4$ &  
$1.1\times 10^7$ & MR \\
& Model 1d  & 960 & 5.0 & 1.9 & 0.31 & 0.75 & $1.3\times 10^4$ & 
$1.1\times 10^7$ & MR \\
&     &&          &     &      & &      & &   \\
\multicolumn{8}{l}{Intermediate cluster models ($kT_{cl}=4\kev=
4.6\times 10^7$K)} &&\\
& Model 2a & 1380 & 1.0 & 1.3 & 24.7 & 0.0 &$5.1\times 10^7$  & 
$5.1\times 10^7$  & ES \\
& Model 2b & 2070 & 1.0 & 2.0 & 65.5 & 0.0 & $4.8\times 10^7$ &
$5.2\times 10^7$  & ES \\
& Model 2c & 1380 & 5.0 & 1.3 & 4.9  & 0.03 & $5.2\times 10^7$ & 
$2.9\times 10^7$ & MR \\
& Model 2d & 2070 & 5.0 & 2.0 & 13.1 & 0.0 & $5.6\times 10^7$ & 
$4.9\times 10^7$ & ES \\
&            &      &     &     &      &    &&  &  \\
\multicolumn{8}{l}{Hot cluster models ($kT_{cl}=8\kev =
9.3\times 10^7$K)} &&\\
& Model 3a  & 2010 & 1.0 & 1.4 & 141.5 & 0.0 & $9.5\times 10^7$ & 
$9.7\times 10^7$ & ES\\
& Model 3b  & 3015 & 1.0 & 2.1 & 374.5 & 0.0 & $9.4\times 10^7$ &
$9.8\times 10^7$  & ES\\
& Model 3c  & 2010 & 5.0 & 1.4 & 28.3  & 0.0 & $9.9\times 10^7$ & 
$9.3\times 10^7$ & ES\\
& Model 3d  & 3015 & 5.0 & 2.1 & 74.9  & 0.0 & $9.8\times 10^7$ & 
$9.8\times 10^7$ & ES\\
\end{tabular}
\begin{flushleft}
Notes on Table~\ref{tab:x3}\\
Column (1): Galaxy velocity with respect to the cluster ICM 
(Section~\ref{sec2p3p2}).\\
Column (2): Galaxy mass-replenishment rate (Section~\ref{sec2p3p3}).\\
Column (3): The Mach number of the galaxy with respect
to the undisturbed ICM.\\ 
Column (4): The ram-pressure stripping
parameter (Section~\ref{sec2p5}).\\ 
Column (5): The galaxy mass-retention fraction (eqn.\,\ref{eq16}).\\
Columns (6) and (7): The mean temperatures in the tail region and galaxy centre
region (Section~\ref{sec2p5}).\\
Column (8): The mode of stripping for this model: ES - \lq\lq Efficient
Stripping\rq\rq; MR - \lq\lq Mass-Retention\rq\rq\ (see Section~\ref{sec3} 
for an explanation).
\end{flushleft}
\label{tab:x3}
\end{table*}

\section{Results}
\label{sec3}

In this section we discuss the basic {\em dynamical} features found in
these simulations. In Section~\ref{sec4} we go on to discuss the expected
{\em observable} features that may be seen at
X-ray energies.

There are two basic dynamical modes of behaviour in these simulations
(illustrated in Figs.~\ref{fig:x1} and \ref{fig:x2}). The
first is when the galaxy is completely stripped by the ICM (and the
mass-retention fraction $\Phi\sim 0$), and there is no substantial build
up of material at the galaxy centre. We term this flow behaviour the
\lq\lq Efficient Stripping\rq\rq\ or ES mode. The second is when there
is a substantial build up of material in the centre of the galaxy, and
the stripping is less efficient ($\Phi\sim 0.1-0.6$). We term this flow
behaviour the \lq\lq Mass-Retention\rq\rq\ or MR mode. In the first case
the dynamics are relatively simple, while for the second case the
dynamics are complex, resulting in a chaotic downstream region. Results
from simulations showing both forms of behaviour are shown in
Figs.~\ref{fig:x1} and \ref{fig:x2}. For each of the model runs listed in
Table~\ref{tab:x3} the main dynamical mode is identified. As discussed
later, however, the distinction between the modes is not always
completely clear cut.

\subsection{The \lq\lq Efficient Stripping\rq\rq\ Mode}
\label{sec3p1}

We illustrate the case of the \lq\lq Efficient Stripping\rq\rq\ mode by
looking, in detail, at two representative simulations, one each for the
intermediate and hot cluster models (Models 2a and 3d). The density and
temperature structure for both models are shown in Fig.~\ref{fig:x1}.

\begin{figure*}
\vspace{6.5cm}
\includegraphics{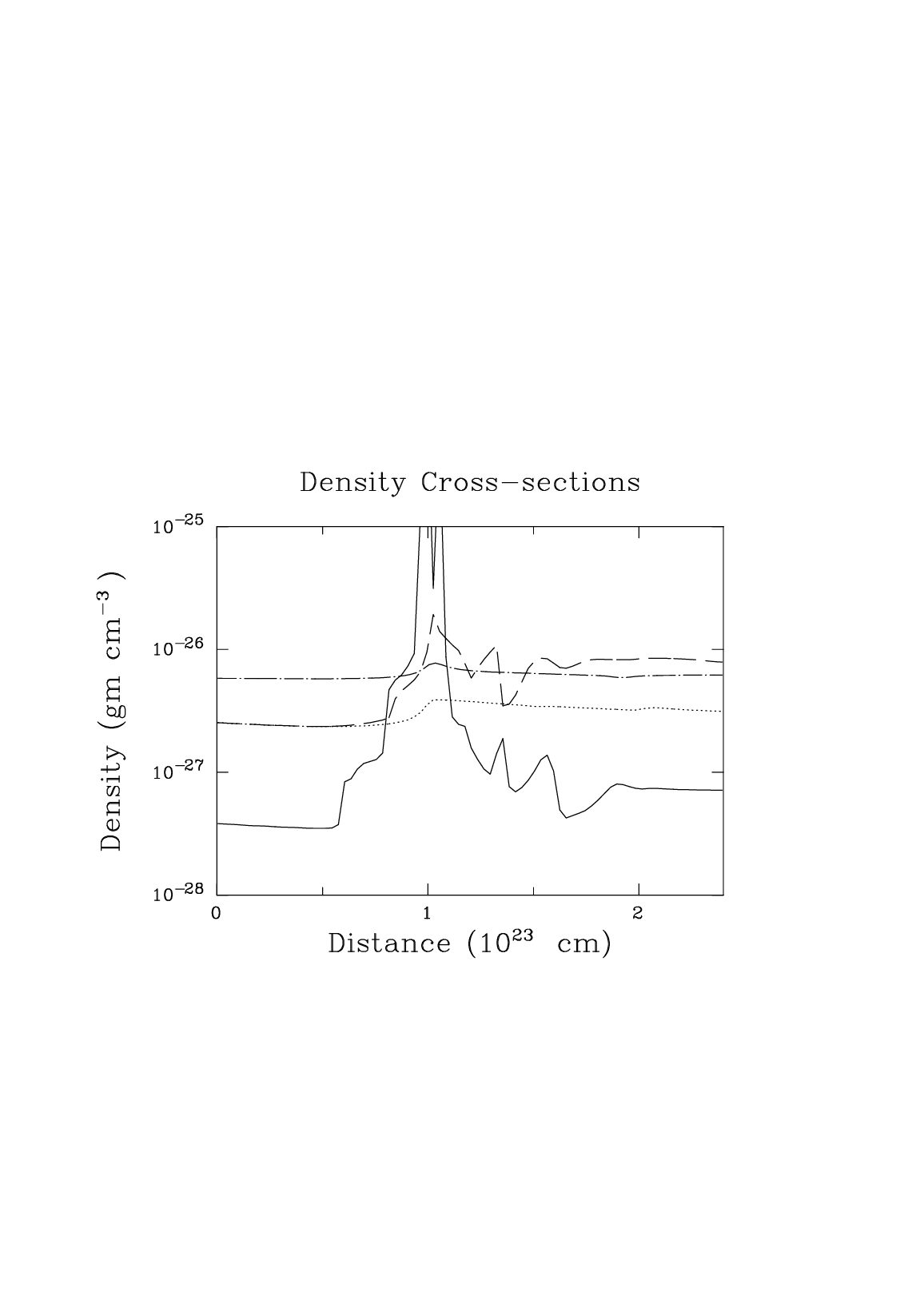}
\includegraphics{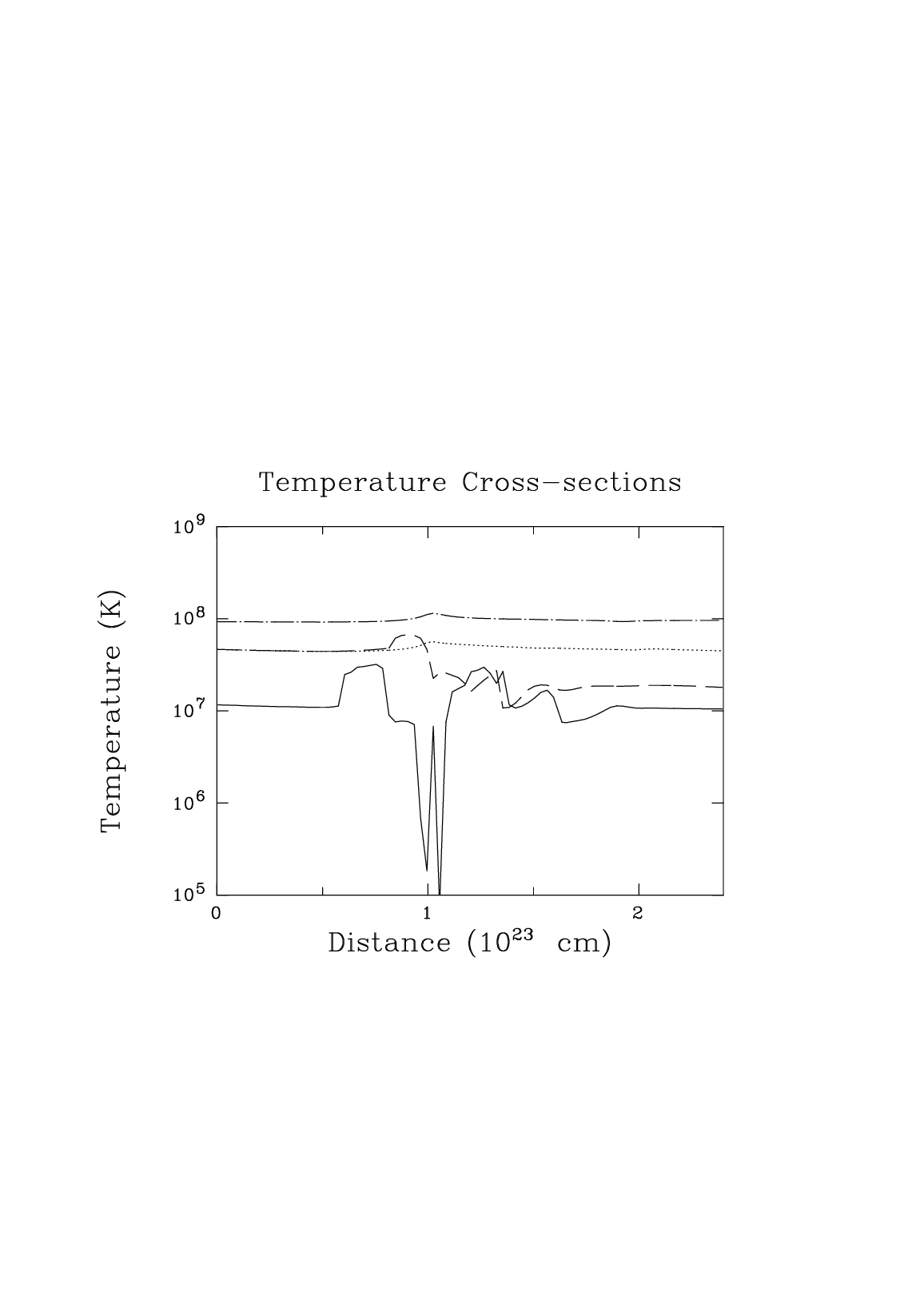}
\caption{Left panel: The density profiles along the $x$-axis for the
four simulations shown in Figs.~\ref{fig:x1} and Fig.~\ref{fig:x2}. Shown
are results for Model~1b (solid line), Model~2a (dotted line), Model~2c
(dashed line) and Model 3d (dot-dashed line). Note, that the cluster
ambient density is function of cluster temperature, with the hottest
cluster having the highest ambient density. There is a clear difference
between these simulations, with ES mode simulations (Models~2a and 3d)
showing little density enhancements, while MR  mode simulations
(Models~1b and 2c) show pronounced density enhancements in both the
galaxy region and the downstream region. In particular, the double shock
structure, discussed by Balsara \etal\ (1994), is  visible for Model~1b
(solid line). Right Panel: The temperature profiles along the
$x$-axis. The lines as the same as in the left panel. The ES mode
simulations show only small temperature perturbations, while  MR mode
simulations show more complex structure.}
\label{fig:x3}
\end{figure*}

Consider Model~2a first. This is a lower Mach number simulation,
with ${\cal M}=1.3$. The density and temperature structure, at a time
$5\times 10^{8}$\,yr into the run, are shown in the top panels of
Fig.~\ref{fig:x1}. The basic flow structure is relatively simple and the
simulation quickly settles down quickly to a steady state. The main
features that can be seen in Fig.~\ref{fig:x1} are as follows:

\begin{enumerate}

\item a prominent bow-shock in front of the galaxy, visible in
both the density and temperature plot. There is an increase in density
and temperature across the shock. The peak post-shock temperature is
$5.7\times 10^{7}$\,K, compared to the ambient ICM temperature of
$4.6\times 10^{7}$\,K (Fig.~\ref{fig:x3}). This increase in temperature
is that given by the Rankine-Hugoniot conditions for a shock transition
with this Mach number.

\item The gravitational acceleration of the ICM material by the galaxy
results in a lower density region preceeding the galaxy.

\item There is little material accumulated at the centre of the galaxy,
with a peak density of $\rho=4\times 10^{-27}\gmcmc$
(Fig.~\ref{fig:x3}). The density never gets high enough for
star-formation to occur. In contrast, the peak density can be much
higher in simulations with a higher $\dot M_{rep}$.

\item The galaxy is completely ram-pressure stripped, with the
mass-retention fraction $\Phi\sim 0$.

\item On account of the relatively small perturbations in density and
temperature the values of $\bar T_{gal}$ and $\bar T_{tail}$ are similar
to the ambient cluster temperature.

\item The ICM velocity field is not greatly affected by the galaxy.
Material in front of the galaxy is gravitationally accelerated, 
then decelerated both by the shock and by the mass-loading of
replenished material, which is swept straight out of the galaxy. 
There are no regions where there is a region of downstream 
reverse flow back in the galaxy (\cf Fig.~\ref{fig:x4}).

\end{enumerate}

The results for Model~3d (shown in the lower panel of Fig.~\ref{fig:x1}),
are essentially the same as for Model 2a. The results are shown at a
time $2.5\times 10^8$\,yr after the start of the simulation. In spite of
the fact that the mass-replenishment rate for this simulation is higher than
for Model~2a the ram-pressure of the cluster is still sufficient to
completely strip the galaxy. From Fig.~\ref{fig:x1}, for Model~3d the
opening angle of the bow-shock is substantially smaller. This is due to
the higher Mach number of the flow (${\cal M}=2.1$) for Model~3d
compared to ${\cal M}=1.3$ for Model~2a.

The expected observational signatures for these simulations will be
discussed in Section~\ref{sec4}.

\subsection{The \lq\lq Mass-Retention\rq\rq\ Mode}
\label{sec3p2}
In contrast to the relatively simple dynamics for the \lq\lq Efficient
Stripping\rq\rq\ cases described above, when the mass-replenishment is 
sufficient or the ram-pressure low enough, complex and time-dependent
flow patterns result in the vicinity of the galaxy. In these cases,
material accumulates in the central region of the galaxy and dense blobs
and streams of material are stripped from the galaxy. In this situation
we do expect to see observable consequences of the galaxy/ICM interaction.

The MR mode is illustrated in Fig.~\ref{fig:x2} with results from two
simulations (Model~1b, top panels and Model~2c, bottom panels), and we
will concentrate on Model~1b initially. The density and
temperature profiles shown in Fig.~\ref{fig:x3} are at a time $6.8\times
10^8$\,yr after the start of the simulation. The MR simulations 
typically do not settle down to a steady-state, as there is a radial
\lq\lq pumping\rq\rq mode instability in the flow (as discovered and
discussed by Balsara \etal\ 1994). However, the MR mode simulations do reach a
broadly quasi-steady state configuration.

As in the case of \lq\lq Efficient Stripping\rq\rq\ a large bow-shock 
region leading and enveloping the galaxy is apparent, 
caused by the interaction of ICM with the galaxy. As noted earlier we
also a leading bow-shock in the case when efficient stripping occurs. In
this case the opening angle of the bow-shock is broadened by the
accumulated replenished material at the centre of the galaxy. Behind the
bow-shock is a region of shock heated gas (though the temperature
enhancements is small given the relatively low Mach number of the flow).

However, the main difference between this simulation and Model~2a
shown in Fig.~\ref{fig:x1} is the complex structure around the
galaxy. There is also a dense region at the galaxy centre, with
ram pressure stripping primarily occurring from the outer
layers of the galaxy. Complex Kelvin-Helmholtz type instabilities
develop at the boundary where stripping is occurring. The
stripping events result in long-dense filamentary structures trailing
the galaxy. A consequence of this stripping is the formation of a dense,
potentially observable, tail (Section~\ref{sec4}).

For Model~1b, within the outer bow-shock is a second shock, resulting
from the inflow from material from the downstream region
passing through the galaxy (and being mass-loaded by the galactic
mass-replenishment) before interacting with the incident ICM. The density
profile for Model~1b is shown in Fig.~\ref{fig:x3}, and shows both shocks.
The formation of this second shock is discussed in Balsara \etal\ (1994).

In Model~1b the ram-pressure stripping is sporadic, with large amounts
of material being stripped over short timescales. These mass-stripping
events occurs on a time scale of a few times the flow-crossing
timescale, $\tau_{flow}$.

\begin{figure}
\vspace{7.7cm}
\includegraphics{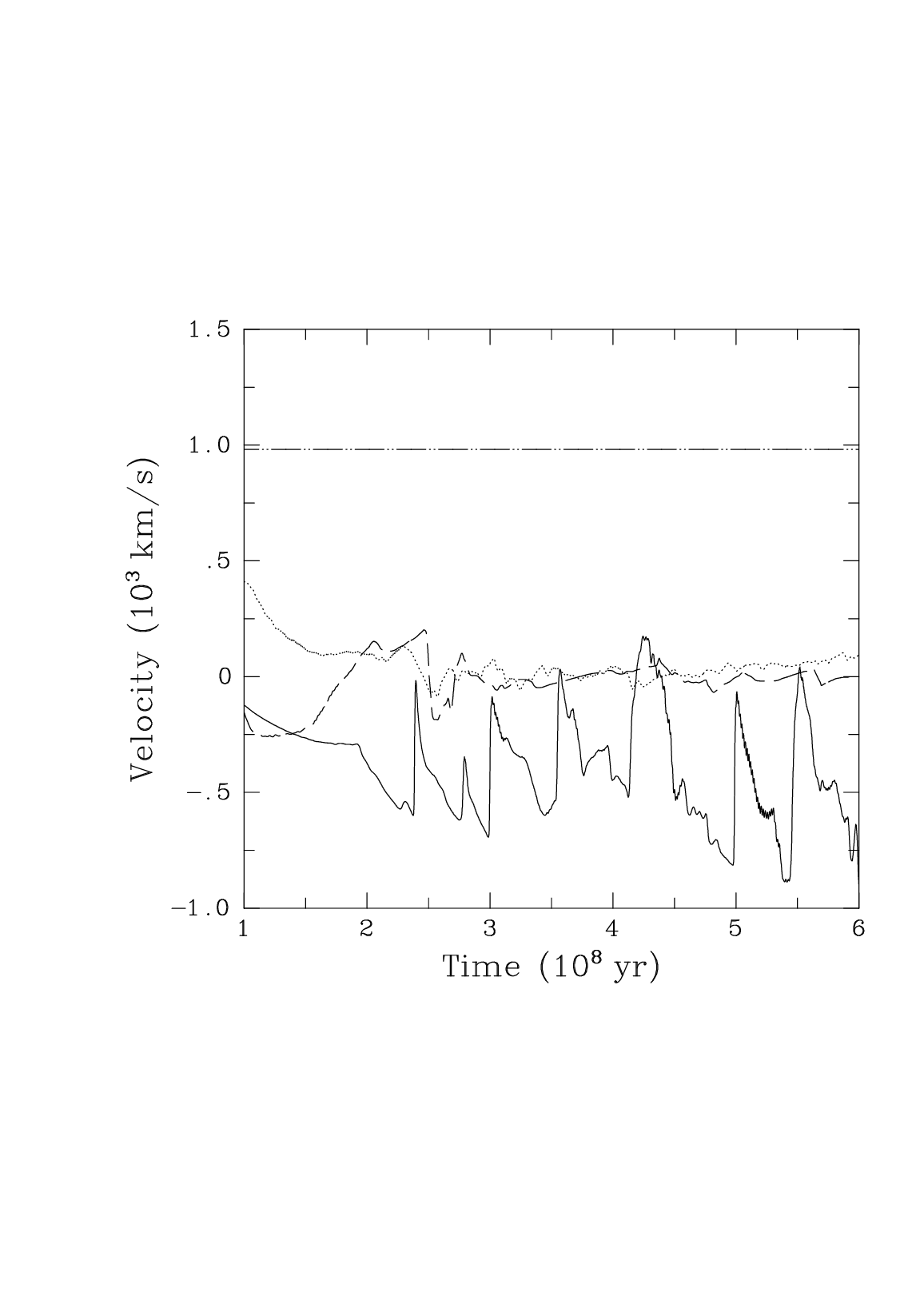}
\caption{The time-variation of the $x$-component of the gas velocity
with respect to the galaxy at several points in the flow for
Model~1b. The sampling positions are at $x=10^{22}$~cm (\ie well in
front of the galaxy - dot-dashed line),  $x=9\times 10^{22}$~cm (\ie a
small distance in front of the galaxy centre - dotted line),
$10^{23}$~cm (at the galaxy centre - dashed line), and $1.5\times
10^{23}$~cm (behind the galaxy - solid line). Behind the galaxy the
$x$-component of the velocity changes sign several times in a
quasi-periodic manner, implying
back-flow into the galaxy, which results in the build-up of material at
the centre of the galaxy. This is the \lq pumping mode\rq\ instability
discussed in Balsara \etal\ (1994).}
\label{fig:x4}
\end{figure}

To further illustrate the complex flow dynamics for Model~1b in
Fig.~\ref{fig:x4} we show the  $x$ component of the velocity at four
different locations in the flow. In the undisturbed region ahead of the
galaxy and bow-shock the gas velocity is essentially constant with time.
At the galaxy centre there is sufficient build up of material to ensure
that any velocities are small. There are, however, velocity excursions
with both positive and negative values. The most interesting flow
pattern is behind the galaxy, where the flow is complex, with
periods of flow into the galaxy from the downstream region. Apparent from
Fig.~\ref{fig:x4} the timescale for these events is $\sim 5\times 10^7$
years, a few times $\tau_{flow}$. This is instability behaviour noted 
by Balsara \etal\ (1994).

For the case of Model~2c, the basic density and
temperature pattern is broadly similar to that for Model~1b
(Fig.~\ref{fig:x3}), in that there is a bow-shock and an enhanced density
tail. However,  Model~2c is a less extreme case than Model~1b, with
the enhancement in density at the centre of the galaxy being only a
factor of a few rather than several orders of magnitude. The tail is
also narrower and shows less complex structure. Model~2c seems to be
close to the 
transition between the ES and
MR mode. At earlier times in the simulation it more closely resembles the
ES case, but as the simulation progressed the velocity at the centre of
the galaxy dropped with progressive mass-loading. Mass  
accumulates at the centre, making it more difficult to strip the galaxy 
(and so on). This simulation illustrates that the distinction between the two
modes of behaviour is not completely clear cut, and in particular as the
environment of the galaxy changes (as it moves closer or further from
the galaxy centre) then the dynamics will also change. So, for example,
while all runs for the hot cluster in Table~\ref{tab:x3} show the ES
mode, even with the high mass-replenishment rate, 
if the galaxy were at a larger radius from the cluster
centre then, because the cluster
ram-pressure is lower, the galaxy would be able to retain much
of the replenished mass, and the dynamics would be very different 
(see Section~\ref{sec3p3}). 
It will also be possible for the flow pattern to undergo
the opposite transition. Consider a situation like Model~1b, with
the galaxy losing mass by discontinuous stripping events. If the 
incident ICM ram-pressure were to increase (either higher density or
higher velocity), as might be expected if a galaxy was falling into the
centre of a galaxy cluster, at some point the ram
pressure would be sufficient to completely strip the galaxy, and the
galaxy would return to the efficient continuous stripping mode.
Calculations by Takeda \etal\ (1984) suggest that this transition can
happen very rapidly, with the expulsion of a dense blobs of gas from the
centre of the galaxy.

\begin{figure}
\vspace{6.5cm}
\includegraphics{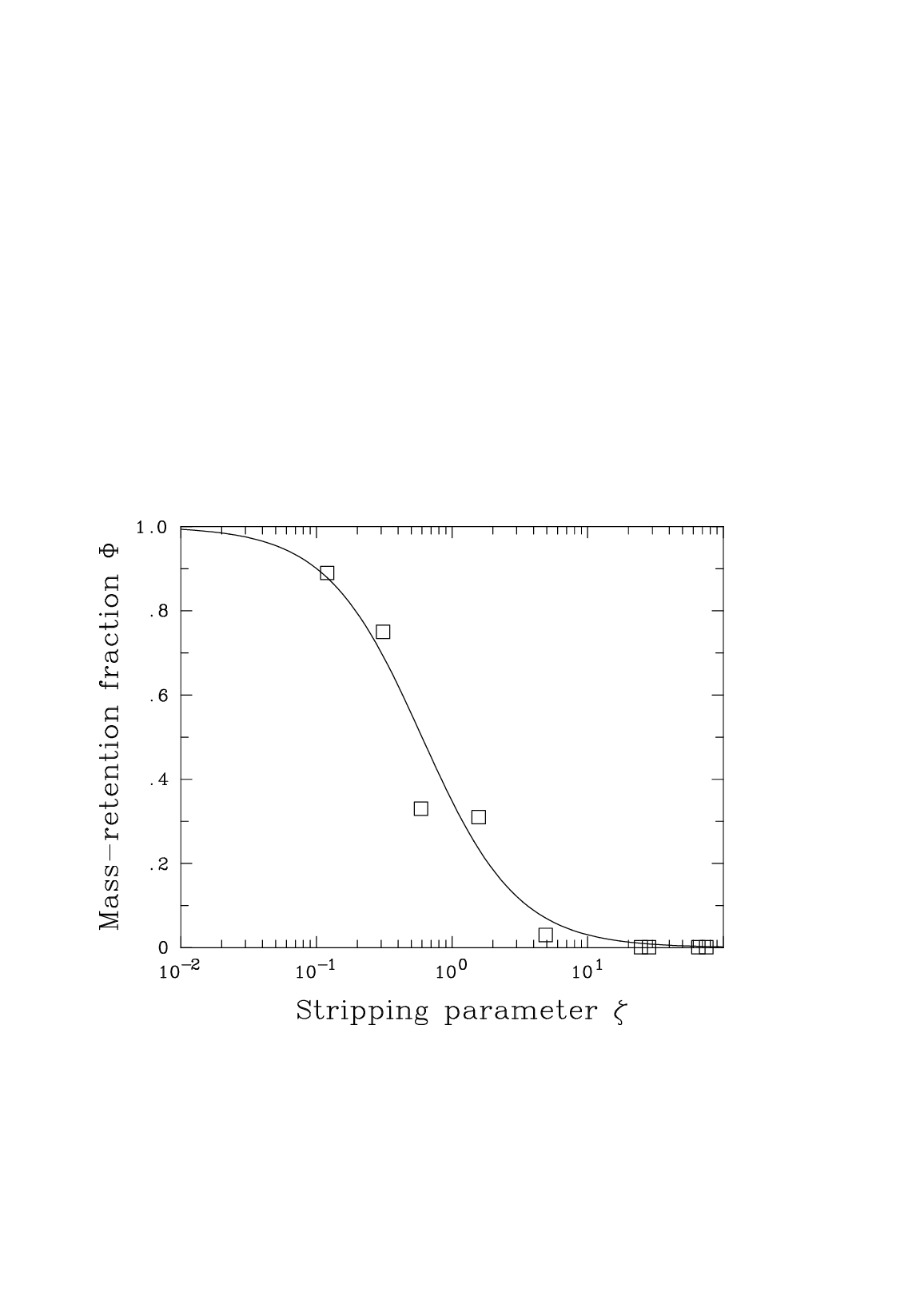}
\caption{The mass retention fraction $\Phi$ versus the mass-stripping
parameter $\xi$. The points are the results from the simulations
presented here and the line is the best fit to the 
results of Gaetz \etal\ (1987; eqn.\,\ref{eq17}).}
\label{fig:x5}
\end{figure}

\subsection{General Considerations}
\label{sec3p3}

Although we have characterised the behaviour of the simulations
presented here into two broad classes, the ES and MR modes, in reality
the distinction is not always clear cut, as illustrated by Model~2c.

In Fig.~\ref{fig:x5} we plot the 
mass-retention fraction ($\Phi$) versus the ram-pressure stripping
parameter ($\xi$) as defined by Gaetz \etal\ (1987). The
best-fit model for the Gaetz \etal\ (1987) simulations is also shown 
(eqn.\,\ref{eq17}). The results of our simulations 
are  similar to those of Gaetz \etal\ (1987), and broadly speaking the
transition from the MR mode to the ES
mode occurs for values of $\xi\sim 5-10$.

Consideration of Fig.~\ref{fig:x2} and the simulations presented in Balsara
\etal\ (1994) shows that for the MR mode material accumulates in the
center of the galaxy, and 
ram-pressure stripping occurs from the
outer edges of this denser central region. 
So, even while a galaxy may retain mass at its
centre, beyond a certain radius it will be completely stripped.
Using the results of Takeda \etal\ (1984) we can evaluate the conditions
from ram-pressure stripping at different radii within the galaxy.

Takeda \etal\ (1984) found that for a spherical galaxy, the ISM
will be stripped from a galaxy at a projected radius $d$ when

\begin{equation}
\rho_{cl} V_{gal}^3 \geq {C\alpha_\ast\Sigma_\ast(d)\sigma_\ast^2}\ , 
\label{eq19}
\end{equation}
where $\rho_{cl}$ is the density of the incident ICM, 
$V_{gal}$ is the galaxy velocity,
$\alpha_\ast$ is the fraction of the stellar mass injected into the
galaxy ISM (\ie for $\dot M_{rep}= 1\msunyr$, $\alpha_\ast=2.6\times
10^{-20}$ s$^{-1}$), $\Sigma_\ast(d)$ is the stellar surface density at a
projected radius $d$, $\sigma_\ast$ the stellar velocity dispersion, and
$C$ a constant to be empirically determined.
For $M_{gal}\sim 10^{12}\msun$ a reasonable value is  
$\sigma_\ast=320\kms$ (\cf Djorgovski \& Davis 1987).

From the stellar distribution used in the hydrodynamic model 
(Section~\ref{sec3}) 

\begin{equation}
\Sigma_\ast(d)=\frac{\pi \rho_\ast(0) R_c^2}{\sqrt{R_c^2+d^2}}\ , 
\end{equation}
with $\rho_\ast(0)$ the galaxy central stellar density. In these simulations a 
value of $C\sim 70$ gives a reasonable representation of the
depth of ram-pressure stripping in the galaxy. For example, for Model~1b
this corresponds to the galaxy being completely stripped beyond a radius
of $d\sim 7R_c$, and for Model~2c for $d\sim R_c$. As discussed by
Gaetz \etal\ (1987), this criterion implies that it is energy transfer
from the incident ICM that is important in the stripping process.
This criterion, although having a  slightly different functionality than
the results from Gaetz \etal\ (1987), can also be used to evaluate the
general behaviour of the simulations. If the equality in
eqn.\,(\ref{eq19}) is satisfied for all $d$ then the galaxy will be
completely stripped. This defines a minimum velocity $V_{strip}$, so
that if $V_{gal}\geq V_{strip}$ then the galaxy will be completely
stripped. On the other hand, if the equality is only satisfied for
$d\geq R_H$ then the galaxy will essentially retain all of the
replenished mass.

\begin{figure}
\vspace{7cm}
\includegraphics{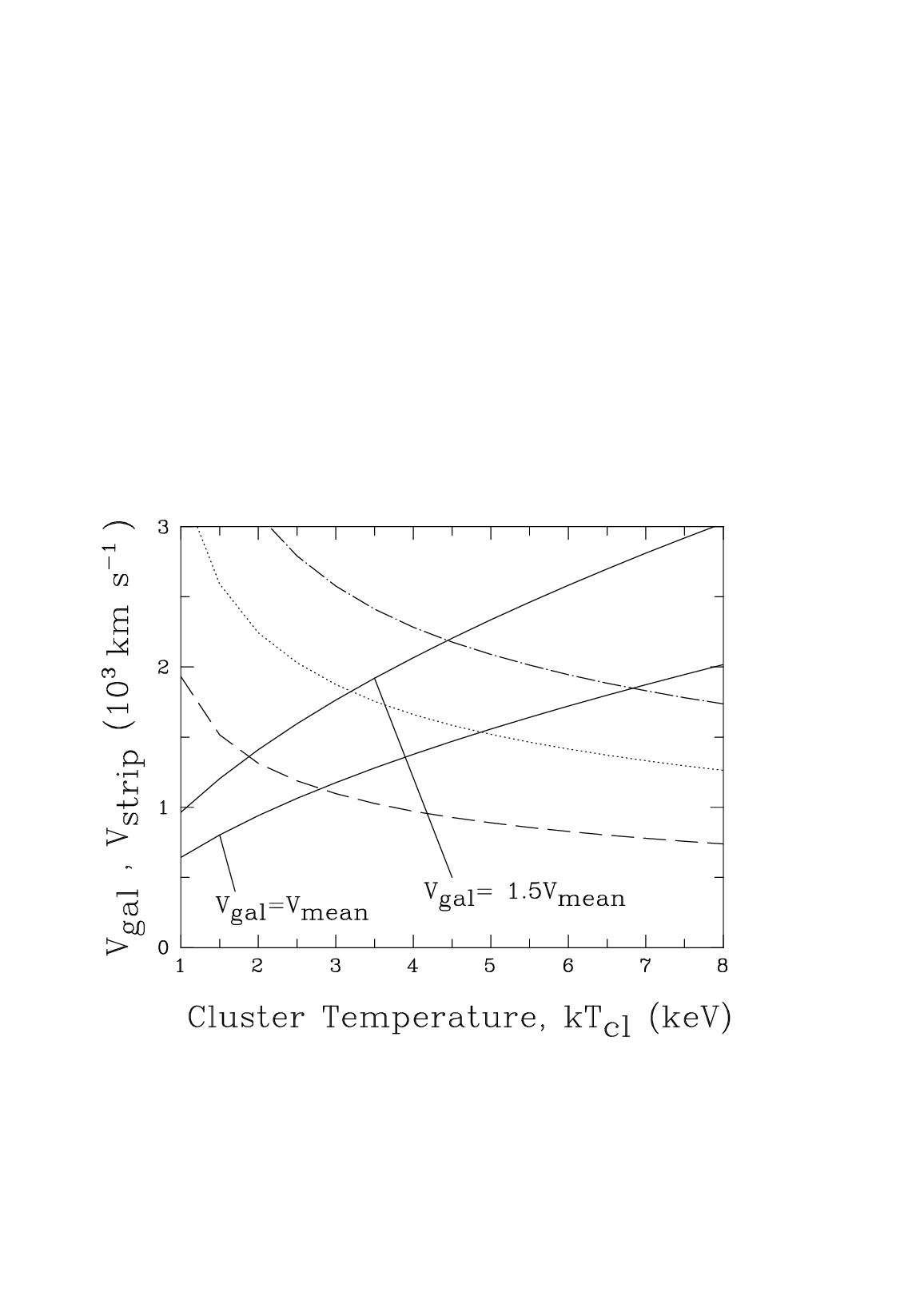}
\caption{The conditions for total ram-pressure stripping from galaxies
as a function of the cluster temperature $T_{cl}$. Shown are the galaxy
velocities $V_{gal}$ as a function of $T_{cl}$ (Section 2.3.2) for the
two assumed conditions $V_{gal}=V_{mean}$ and $V_{gal}=1.5 V_{mean}$
(solid lines). The value of $V_{strip}$ for the case of
$\dot M_{rep}=1\msunyr$ is shown (dashed line), as well as the cases
when $\dot M_{rep}=5\msunyr$ (dotted line). Also shown is the case
of $\dot M_{rep}=1\msunyr$, but with the
galaxy at $r=5R_{core}$ from the cluster center instead of 
$r=R_{core}$ (dot-dashed line). For complete ram-pressure stripping 
$V_{gal}>V_{strip}$. (see Section 3.3 for more details.}
\label{fig:x6}
\end{figure}

In Fig.~\ref{fig:x6} we show the conditions for total stripping based on
the parameters used in this paper. The relationship between $V_{gal}$
and $kT_{cl}$ is shown for the cases when $V_{gal}=V_{mean}$ and 
$V_{gal}=1.5V_{mean}$. 
Also shown is $V_{strip}$ as a function of
$kT_{cl}$. Essentially, $V_{strip}\propto T_{cl}^{-0.5}$. The case when
the galaxy is at a radial distance of $r=R_{core}$ from the cluster
centre and $\dot M_{rep}=1\msunyr$ is shown (dashed line), and suggests
that for a cluster with a temperature above $3\kev$\ the galaxy will be
completely stripped (for $V_{gal}=V_{mean}$). 
For a larger mass-replenishment rate of 
$\dot M_{rep}=5\msunyr$ the required temperature for complete stripping 
rises to $\sim 5\kev$\ (dotted line). In the simulations presented here,
for the $kT_{cl}=4\kev$\ cluster and for $\dot M_{rep}=1\msunyr$ the
galaxy is completely ram-pressure stripped, while for $\dot M_{rep}=5\msunyr$
the galaxy is stripped for the larger value of $V_{gal}$ but not for the
smaller value. Moving the galaxy out to larger
distances from the cluster centre increases the stripping temperature
(for large $r$ $V_{strip}\propto r^{2/3}$),
and means that even galaxies in hot clusters will not necessarily
be completely stripped if they are on the periphery of the cluster
(dot-dashed line). It is worth noting that for the Fornax and Virgo clusters 
(with $kT\sim 1.3\kev$ and $1.9\kev$ respectively) that for 
$\dot M_{rep}=1\msunyr$ then $V_{strip}=1680\kms$ and $1350\kms$
respectively. 

The main point from Fig.~\ref{fig:x6} is that the formation
of dense tails behind galaxies in clusters is most likely to occur in
cool clusters, for galaxies with high mass-replenishment rates, or
alternatively in the outer regions of richer clusters.

We also note the following general trends from the simulations. Gas directly
behind the bow-shock is always hotter than ambient cluster. On the other
hand, gas from ram-pressure stripping will be cooler than the ambient cluster
gas ($T_{rep}\ll T_{cl}$). Depending
on the balance between the two sources of material the tail will be
either hotter than (or comparable to) the cluster or cooler than the
cluster if substantial stripping is occurring. This process is illustrated
in the values of $\bar T_{tail}$ in Table~\ref{tab:x3}, where for 
MR mode simulations $\bar T_{tail}\ll T_{cl}$, while for the ES 
mode $\bar T_{tail}\sim T_{cl}$.

One initially counter-intuitive result that follows from this is that
those galaxies which retain a larger fraction of gas (and consequently
suffer less from ram-pressure stripping) have denser tails, which in
turn translates into more visible tails. Correspondingly, galaxies 
that lose the most material into the
downstream region have less dense tails. The reason is 
straightforward - the density (and visibility) of the downstream region
is a sensitive function of the gas-dynamics behind the galaxy. Those
galaxies that have higher values of $\dot M_{rep}$ have more complex
flow patterns behind the galaxy, with back-flow occurring and generally
lower velocities in the tail. This should be contrasted with low
mass-replenishment cases where the flow is simple - there is some
deceleration associated with the galaxy, but the $x$-component of the
velocity is always positive (\cf Fig.~\ref{fig:x4}). This means that
material stripped from the galaxy is moved swiftly away from the galaxy
and the density never builds up to significant levels.

As might be expected there is a correlation between the opening angle of the 
bow-shock cone and the Mach number. For these simulations we find that the 
opening angle for Mach numbers of ${\cal M}\sim 1.3$ is around $50^\circ$, 
decreasing to around $30^\circ$ for Mach numbers around 2.
While in theory this would allow a method for determining the Mach 
number of the galaxy through the ICM (and hence the actual 3-D velocity 
of the galaxy) there are some complications. If there is 
sufficient mass-loss from the galaxy then this will alter the opening angle. 
In cases where the mass-replenishment is substantial, 
the cone opening angle is broader, which would result 
in an underestimate of the
galaxy Mach number. As discussed later, the bow-shock will usually have 
a faint signature and may be difficult to observe.

\begin{figure*}
\vspace{11.5cm}
\includegraphics{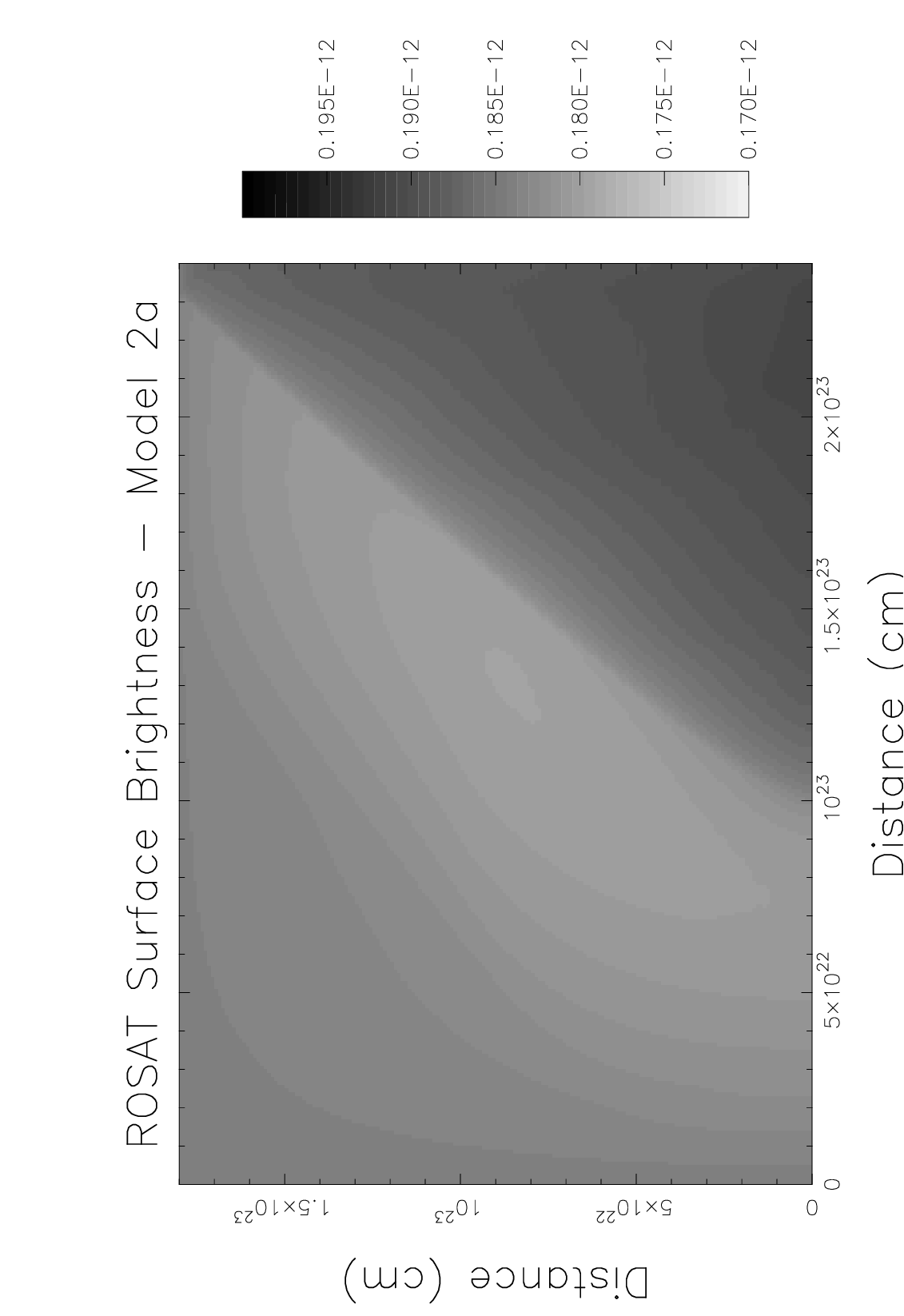}
\includegraphics{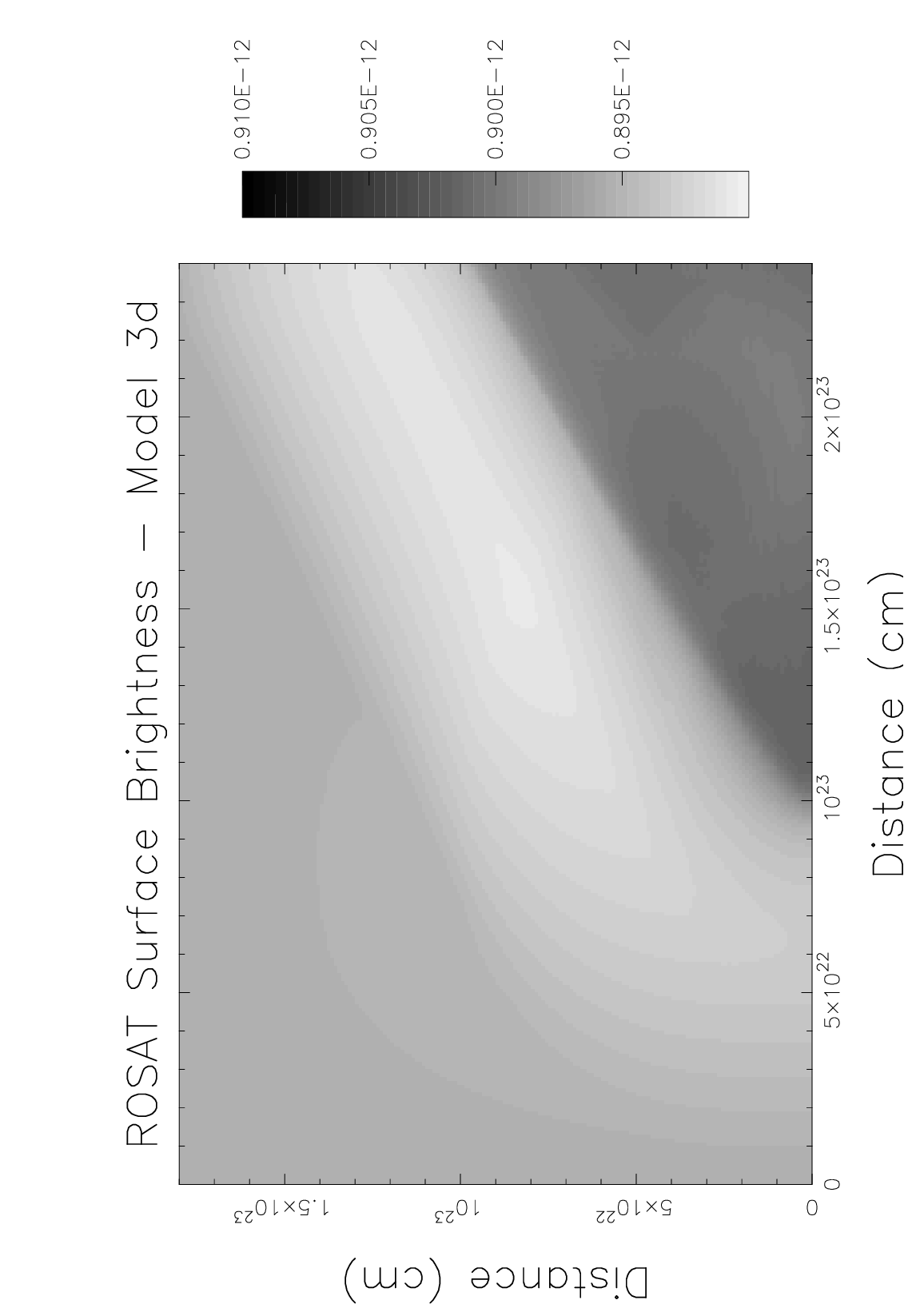}
\includegraphics{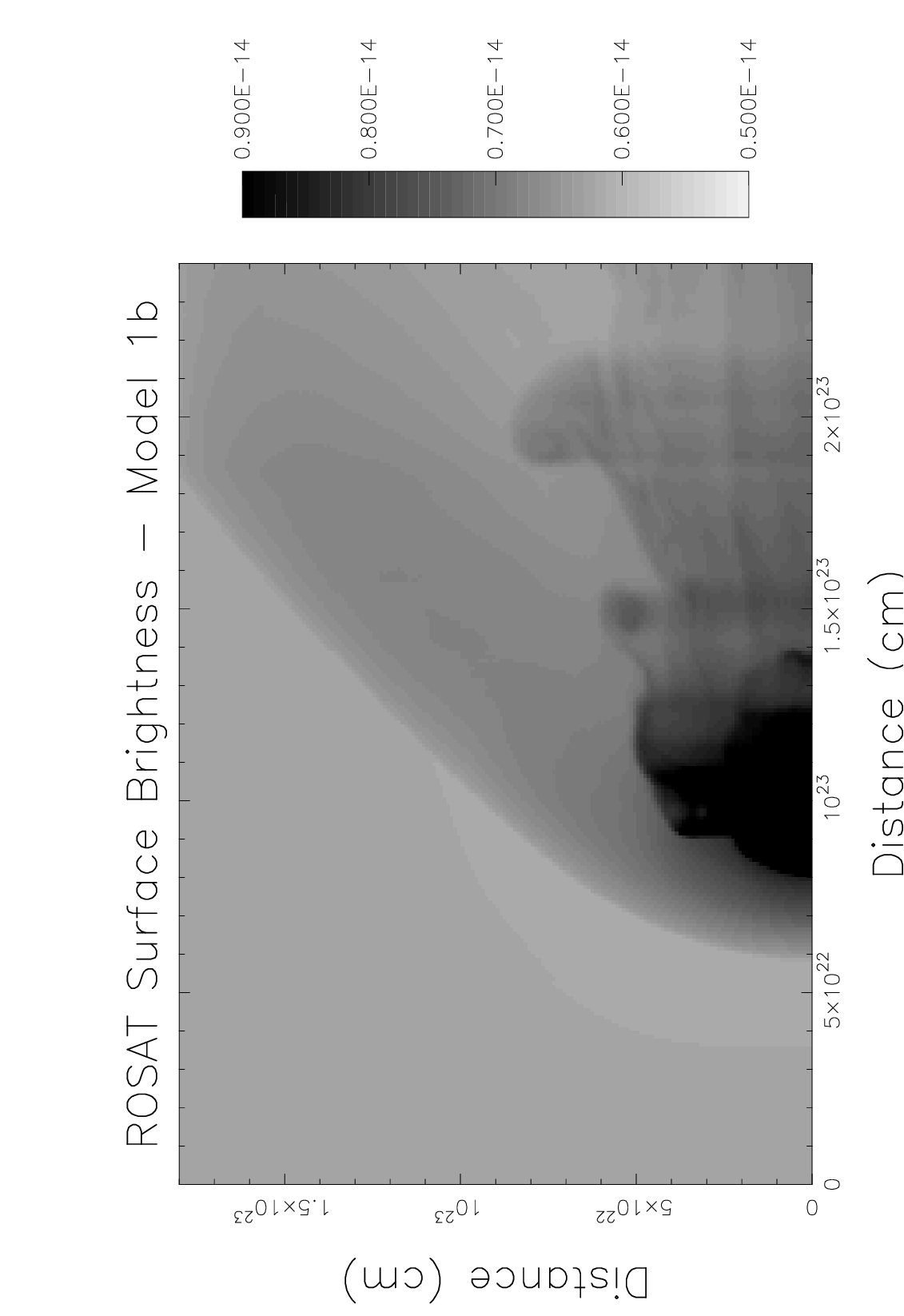}
\includegraphics{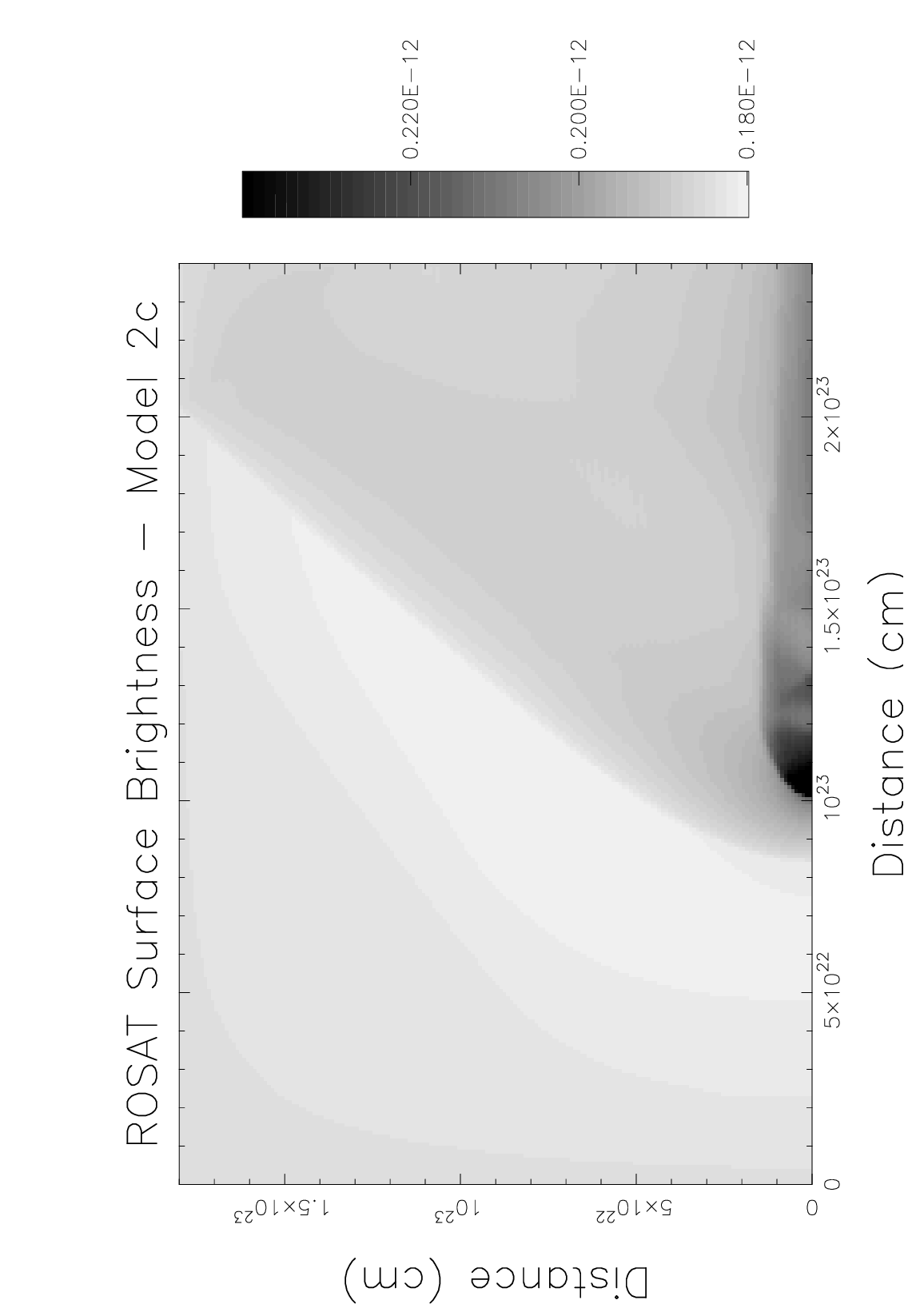}
\caption{Synthetic \rosat\ surface brightness profiles
for the four simulations shown in Figs.~\ref{fig:x1} and\ref{fig:x2}. Top left
panel: Model~2a; top right panel: Model~3d (both ES mode).
Bottom left panel: Model~1b; bottom right panel: Model~2c (both MR
mode). The synthetic surface brightness images have been calculated in
the $0.1-2.5\kev$\ waveband, with units of \surfb.
Note that emission from the cluster has been included. 
Careful attention should be paid to the
scaling of these images, as some of them have a much larger dynamic
range than others.}
\label{fig:x7}
\end{figure*}

\section{Observable Consequences}
\label{sec4}

One of the main purposes of this  paper is to investigate the expected 
{\em observable} properties of the theoretical simulations of the galaxy/ICM
interaction, with an emphasis on the X-ray waveband. As most of the
current data have been taken with the \rosat\ satellite (the
$0.1-2.5\kev$\ waveband) this is the spectral region we concentrate on,
though future work will look at both the \xmm\ and \axaf\ wavebands. 

\begin{figure*}
\vspace{11.5cm}
\includegraphics{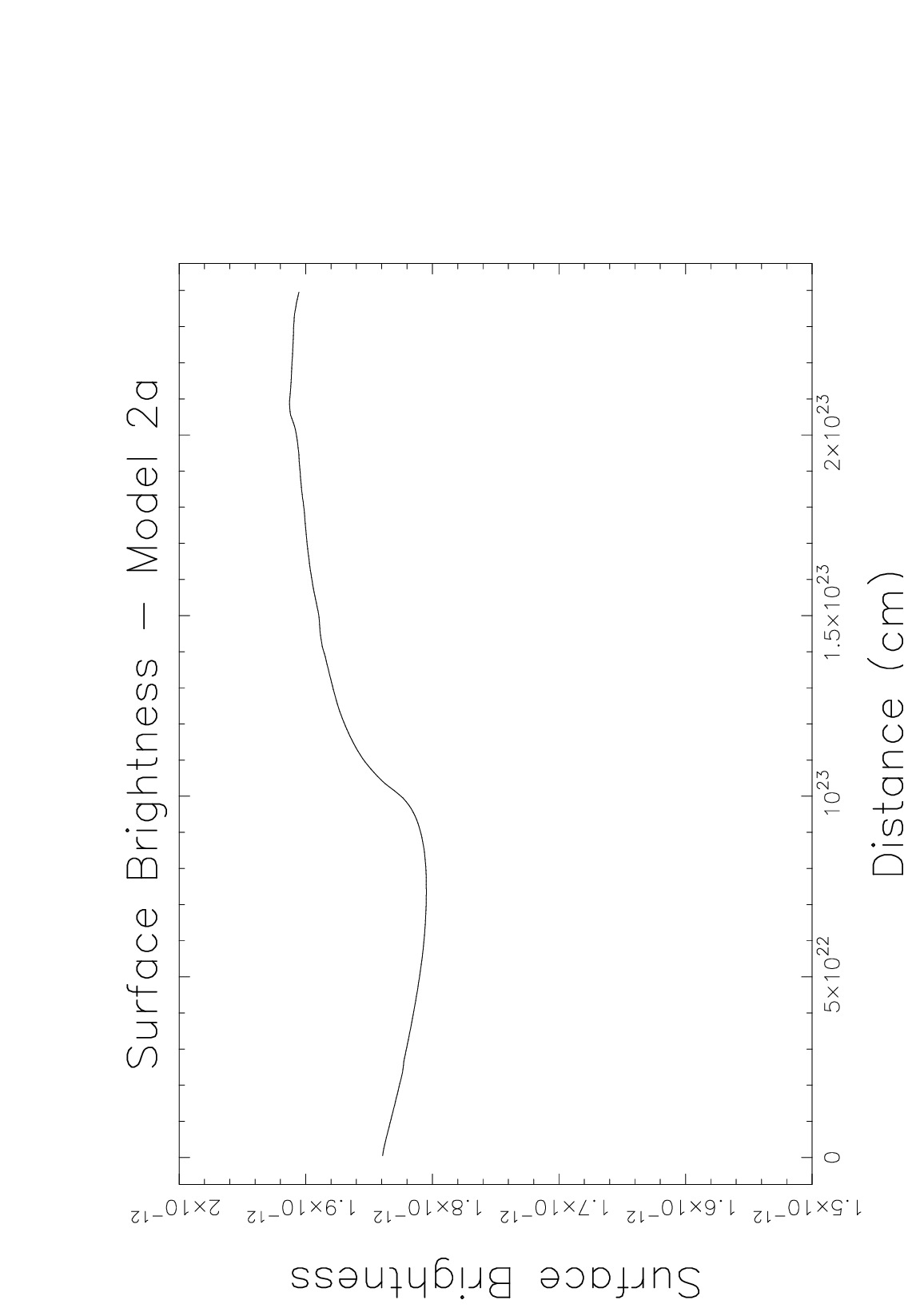}
\includegraphics{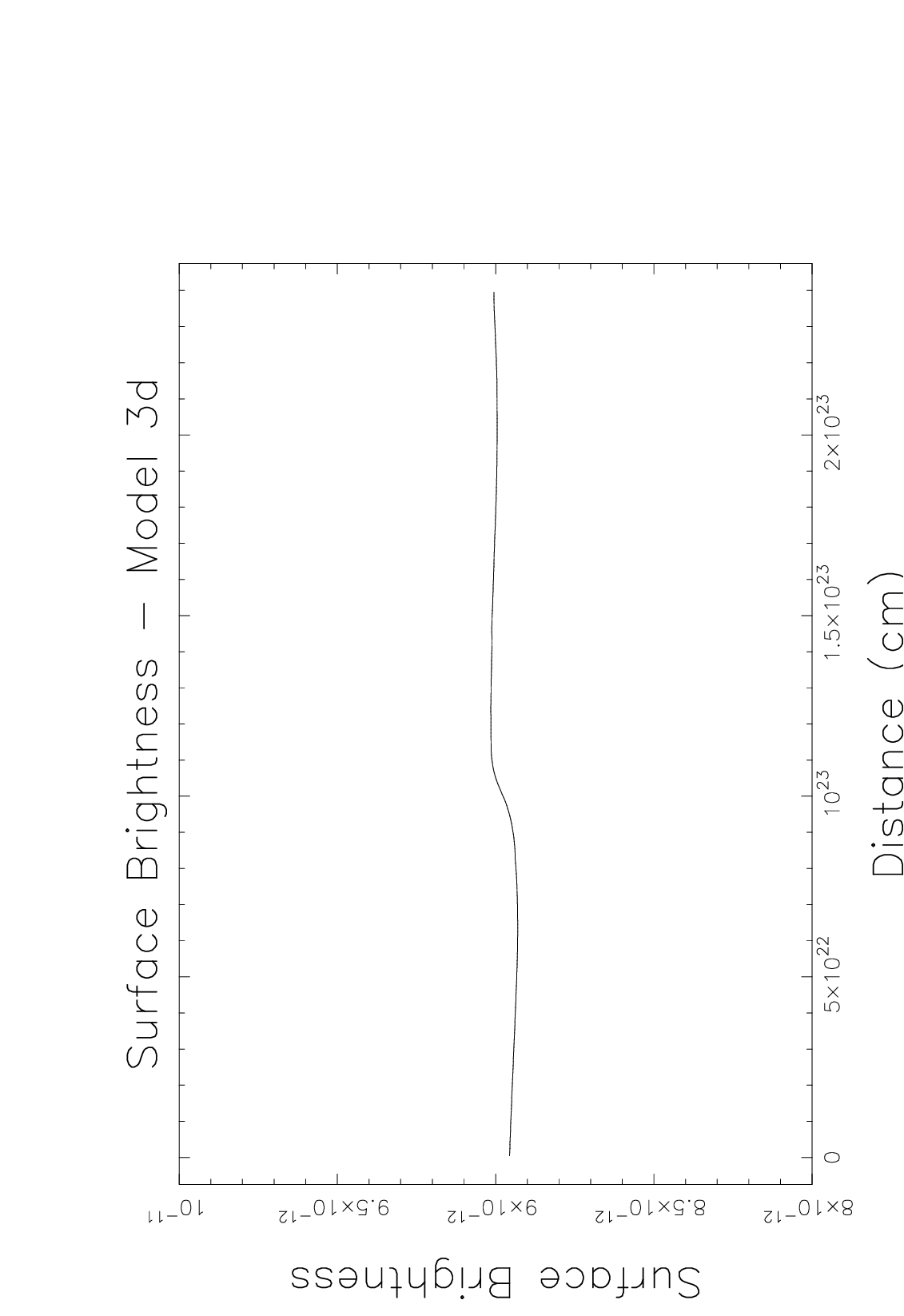}
\includegraphics{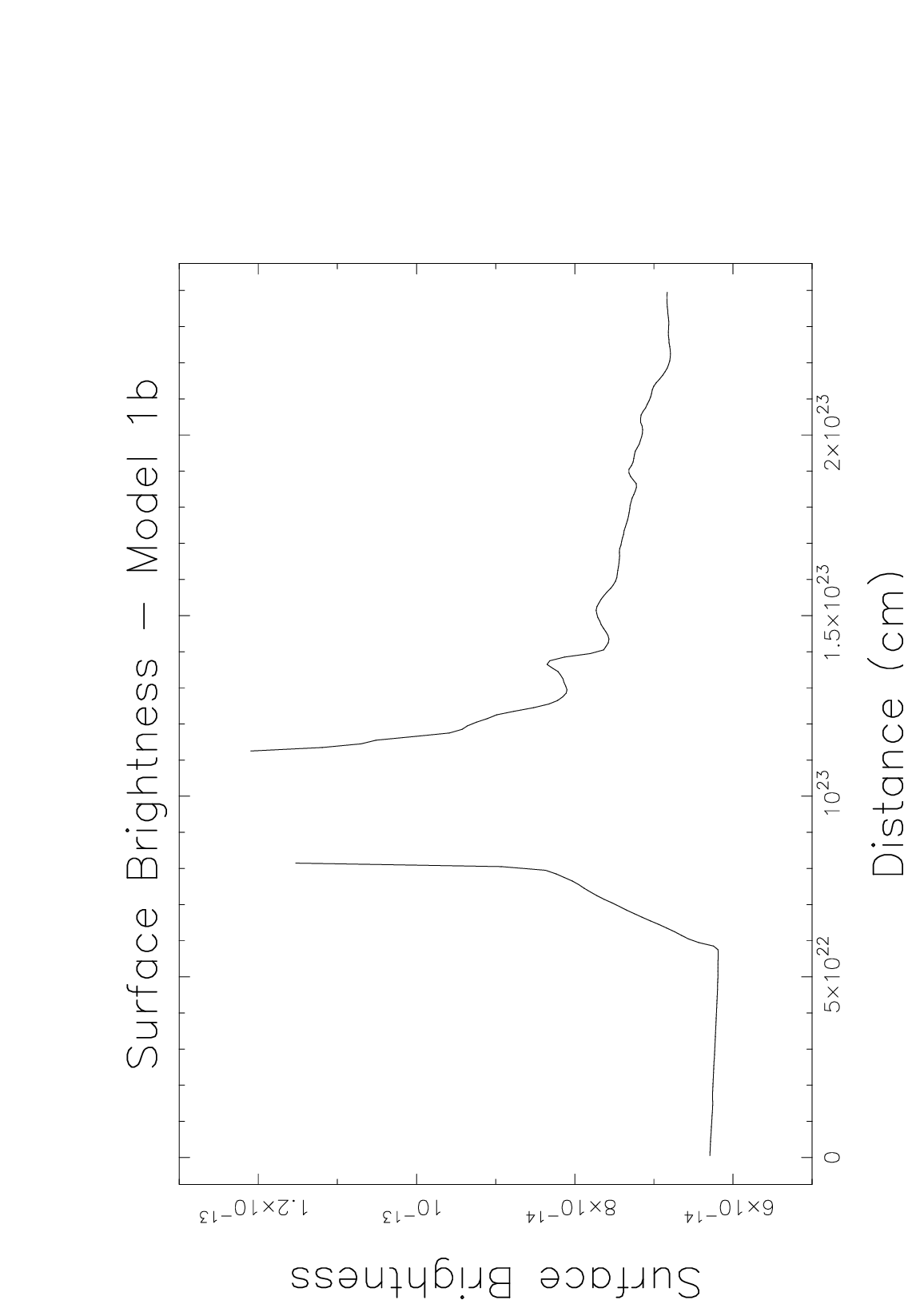}
\includegraphics{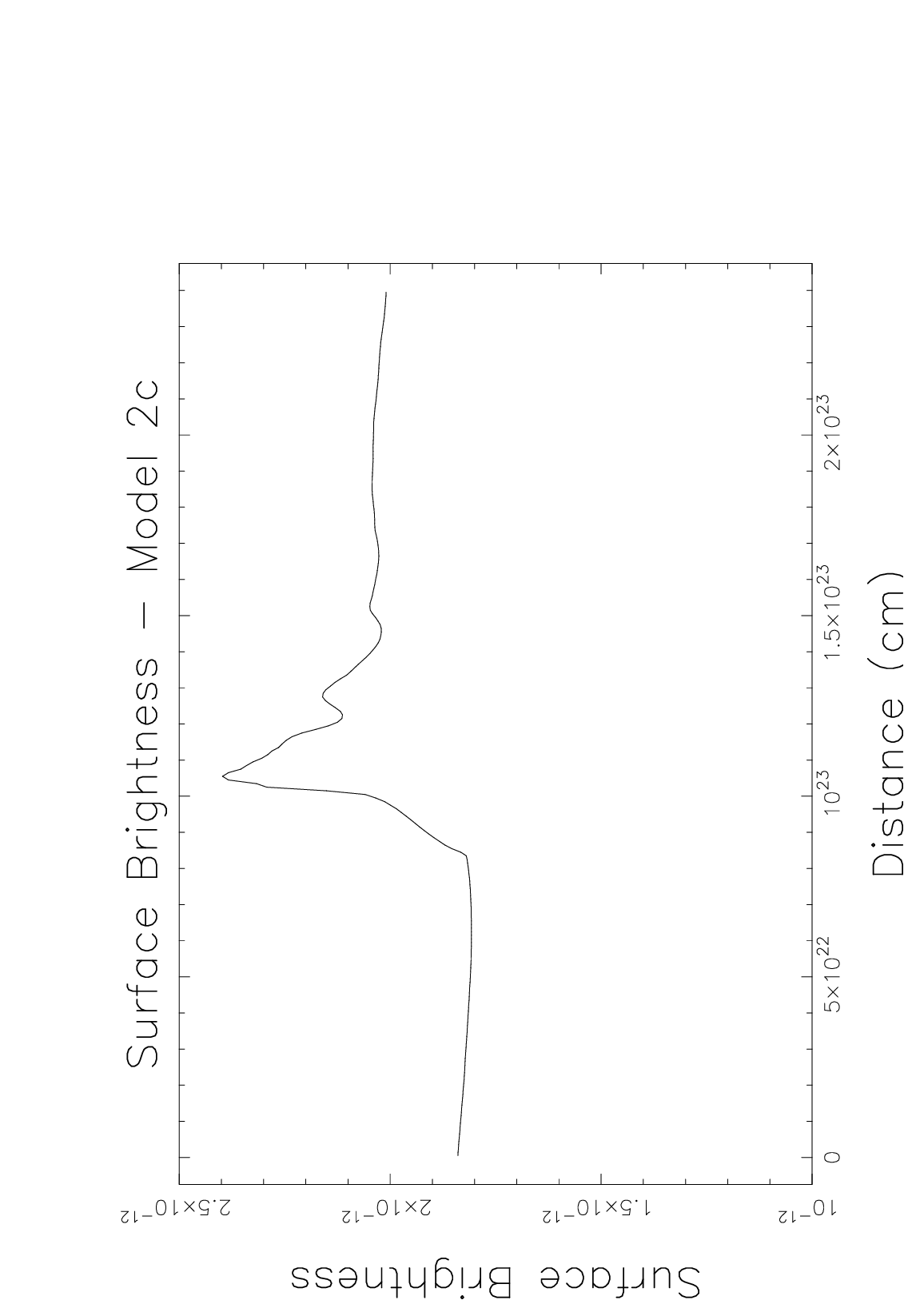}
\caption{Surface brightness slices for the four images shown in
Fig.~\ref{fig:x7}. The units are \surfb. These slices in surface
brightness are taken along the $x$-axis through the galaxy for a slice
with a width of $10^{22}$\,cm. The level of background cluster 
emission varies substantially from the cool to the hot cluster. 
Also note that the scale for Model~1b has been truncated to enhance
the contrast in the tail region. The peak surface brightness for this
model is $\sim 10^{-11}\surfb$. Note the substructure in the tails of
the MR mode simulations.}
\label{fig:x8}
\end{figure*}

The hydrodynamic simulations (Figs.~\ref{fig:x1} and \ref{fig:x2})
provide values of $\rho$ and $T$ at all locations, and then it is
straightforward to generate synthetic X-ray surface brightness maps for
the simulations. The basic synthetic X-ray surface brightness maps for
the four simulations in Figs.~\ref{fig:x1} and \ref{fig:x2} are shown in
Fig.~\ref{fig:x7}. These images are for the $0.1-2.5\kev$\ waveband, and 
the background cluster emission (appropriate for the waveband and for
the cluster surface brightness - Section~\ref{sec2p2}) has been added. 
The cluster emission is often greater than the emission from
the wake or bow-shock region and this has the effect of making the contrast
between the tail or wake and background emission small. 
The effects of Galactic absorption are not included in
Fig.~\ref{fig:x7} as they are likely to be of little consequence.
It is necessary to carefully look at the scales in Fig.~\ref{fig:x7} to
determine the visibility of the various structures. The images can
either have a large (\ie Model~1b) or small (the ES mode simulations) dynamic
range, and the scaling has been adjusted appropriately.
The important features to notice in Fig.~\ref{fig:x7} are as follows:

\begin{enumerate}
\item For the ES mode simulations the most prominent structures are
the bow-shocks, with a small deficit of emission preceding the bow
shock. This is due to the lower density region caused by the
gravitational acceleration of material by the galaxy. 

\item For the hotter cluster simulations (see Model~3d), the cluster
background is very bright, and the contrast between bow-shock and
background is small ($\sim 1$ per cent). This is more apparent in
Fig.~\ref{fig:x8}, where surface brightness slices are shown for all 4
simulations. This means that interaction effects will be more
likely to be seen in cooler, less luminous clusters.

\item For the MR mode (Models~1b and 2c) there are three main features
visible in the synthetic images. First, the bow-shock, second elongated 
emission from the tail region and third emission from the galaxy itself. 

\item For Model~1b, the galaxy is very luminous, and in order to enhance
the visibility of the tail and bow-shock region we do not show the full
dynamic range (\cf  Fig.~\ref{fig:x8}). There is a large accumulation of
material at the galaxy centre. However, this emission, essentially 
analogous to a cluster cooling flow, will likely be strongly absorbed in
a real galaxy. 

\item For Model~2c, both the bow-shock and tail are visible, but the
tail is less pronounced that for Model~1b.

\item For both Model~1b and Model~2c the emission from the tail
does stand out substantially from the cluster background. However, the
contrast for the bow-shock is less. This seems to be a general
characteristic - the bow-shock is always less visible than any
ram-pressure stripped tail. 

\item For Model~1b the synthetic image shows
considerable substructure on a scale of around $5-10{''}$. Such
structures should be clearly visible with \axaf\ and \xmm.

\end{enumerate}

To further illustrate these points, in Fig.~\ref{fig:x8} surface
brightness slices are shown. These slices
have been generated by taking a slice of width $10^{22}$\,cm along 
the $x$-axis. For the ES cases (Models~2a and 3d as examples) the increase in 
surface brightness is relatively small
(a few per cent at most) and the surface brightness 
profile is relatively simple.
For the MR mode case  there are much greater increases in
emission. Taking Model~1b as an example, the region associated with the
galaxy is extremely bright, but the tail region also shows a 
substantial increase, with an excess brightness over the background of 
$\sim 10 -30$ per cent. Even the case of Model~2c, which shows less in
the way of enhanced emission at the position of the galaxy, has a tail
for which the surface brightness is more than the $\sim 10$ per cent
greater than the underlying cluster.

For both the MR mode simulations in Fig.~\ref{fig:x8}, there is
substantial substructure in the surface brightness in the tail region,
with regions of enhanced emission associated with blobs of gas stripped
from the galaxy. The bow-shock region in the MR mode cases extends
further in front of the galaxy than for the ES mode cases.

For our simulations we have generated hardness ratio maps, with 
the hardness ratio defined as $(H-S)/(H+S)$, where $H$ is the hard-band
($0.5-2.5\kev$) image and $S$ the soft-band image ($0.1-0.5\kev$).
The results in Table~\ref{tab:x3} show that for the MR mode the tail
region is cooler than the ambient cluster, and consequently 
spatial hardness variations might be expected. However, because the
temperature differences are usually small and  
the emission from the tail region is always small compared to the
background cluster emission, this means that differences in the hardness
ratio are always small.

\section{Discussion}
\label{sec5}

In this paper we have presented some new simulations of the interaction
between galaxies and the ICM in clusters. For the first time we have
attempted to quantify the likely observational characteristics 
of structures resulting
from such interactions, with some promising results. Galaxies which are
capable of retaining some mass and have
substantial tails are likely to have features that are
observable. However, those that are completely ram-pressure stripped are
unlikely to have observable features. The bow-shock is a more prominent
feature in MR mode galaxies than in ES mode galaxies
(Figs.~\ref{fig:x7} and \ref{fig:x8}).

The likely criterion for galaxies to show observable interaction
effects in clusters are as follows:

\begin{enumerate}
\item Galaxies with younger stellar populations (and hence more
substantial mass-loss) are more likely to show interaction
effects at X-ray energies. 

\item Galaxies in cooler clusters or group will be more likely to have
observable stripped tails. Galaxies in rich clusters will be more
likely to be completely stripped.

\item Galaxies in lower surface brightness clusters - in richer,
brighter clusters the contrast between tail and background emission is smaller.

\item Galaxies at the periphery of rich clusters. Massive gas-rich
galaxies infalling into rich clusters should show tails.
\end{enumerate}

This means that bluer galaxies in nearby cooler clusters are the most
likely place to look for such interaction effects. For rich clusters,
such as Coma, most galaxies are likely to be completely stripped, except
those at the periphery of the cluster, which may be in an environment
where they can retain some mass. These simulations have also shown that
while tails may be a feature of galaxies in clusters of all levels 
of richness the effects are more likely to be seen in the cooler clusters.

The potential rewards of being able to exploit the galaxy/ICM
interactions are great (Merrifield 1998), however these
simulations show the dynamics are complicated and the level of
emission from the tail (for instance) will depend on parameters that 
may be difficult to determine (such as $\dot M_{rep}$). Consequently, it
may take a substantial effort at modelling individual galaxies to fully
determine the true velocity vector for that galaxy. The other, simple
approach, to merely use the galaxies direction of motion on the plane of
the sky (as revealed by a linear trailing structure), may also be able
to yield interesting results.

The assumptions made here have been chosen to be simple and to be the
most favourable. For example, we have assumed the galaxies to be
spherical. In reality, the galaxies will be elliptical and the gas flow
out of the galaxy will depend on the orientation of the galaxy with
respect to its velocity vector through the ICM. For instance, it might
be anticipated that mass might be lost more easily from the galaxy
through the minor axis of the galaxy. We have also assumed that we are
viewing the galaxy in the plane of the sky. If the galaxy is moving at
a different angle then projection effects will play a role,
foreshortening the tail and potentially making identification of
galaxy/ICM features difficult.
The effect of both of these assumptions will have to be explored with
more sophisticated simulations. While certain orientations of an
elliptical galaxy can be dealt with by 2-D simulations, the more
general case of random galaxy orientation will require 3-D simulations

For several of the suspected examples of galaxy/ICM interaction
(Section~\ref{sec1}) all that can be discerned about the dynamic
structures associated with the galaxy motion is an extension of the
X-ray contours away from the cluster centre (usually interpreted as due
to infall of the galaxy towards the cluster centre). Examples of this
are NGC\,4839 in the Coma cluster (Dow \& White 1995) and NGC\,1404
(Jones \etal\ 1997). It is notable that NGC\,4839 is located towards the
edge of the Coma cluster, where we might expect interactions effects to 
be more visible.

In the case of M86, in addition to an emission peak at the galaxy
centre, there is also a second peak associated with the tail. These
simulations and those of Balsara \etal\ (1994) suggest that stripping
will often occur in discrete events, with blobs of material being
removed. These blobs will result in enhancements in the tail region,
and this is what may be occurring in M86. 

NGC\,4472 shows very complex emission, with some evidence of a bow-shock
as well as a tail. This galaxy will clearly require more detailed work
on it. Irwin \& Sarazin (1996) suggest that some of the features may be
related to the elliptical nature of the galaxy and its orientation with
respect to its motion. This idea will clearly have to be tested with
further simulations allowing for non-spherical galaxies.

In summary, we have new presented calculations of the galaxy/ICM
interaction in clusters, with particular emphasis on investigating the
likely observable consequences of such interactions. From these
simulations it is clear that many galaxies should show observable
effects of the galaxy/ICM interaction, the most prominent feature being
linear tails, resulting from ram-pressure stripping. We have performed
simulations for a wide range of cluster parameters, with the goal of
trying to define the best parameter space to investigate. While massive
galaxies, with young stellar populations, in cool nearby clusters such
as Virgo or Fornax offer the best possibilities, the periphery of
richer clusters also offers interesting possibilities. An important
result is that it is not galaxies that are completely ram-pressure
stripped that have the brightest tails, but rather those which are able
to retain a substantial fraction (but not all) of the replenished
material. In this situation, stripping occurs in discrete events and
results in high density tails. Bow-shocks should also be visible in some
galaxies, particularly in
upcoming \xmm\ and \axaf\ observations of nearby candidates. There are 
good prospects that, through a combination of modelling and
observations with the new X-ray observatories, galaxy/ICM
interactions offer a new window to investigating cluster dynamics.
 
\section*{Acknowledgements} 

IRS is supported by a PPARC Advanced Fellowship. DMA was supported by
funding from the School of Physics and Astronomy at Birmingham
University. The calculations and analysis presented here were done on the
{\em Starlink} node at the University of Birmingham.

\end{document}